\documentclass[a4paper,10pt]{elsarticle}

\usepackage{mathrsfs,amsmath,amssymb,breqn}
\usepackage[numbers]{natbib}

\biboptions{numbers,sort&compress}
\usepackage{graphicx}
\usepackage{epstopdf}
\usepackage{graphics}
\usepackage[T1]{fontenc}
\usepackage{ae,aecompl}
\usepackage{times}

\journal{Physica A}

\begin{document}

\title{A non-local quasi-equilibrium state in the Bhatnagar-Gross-Krook Boltzmann equation for thermo-hydrodynamics: Conservation laws, the Boltzmann H-theorem, and the fluctuation-dissipation theorem}


\author{Hiroshi Otomo}\corref{mycorrespondingauthor}\ead{hiroshi.otomo@tufts.edu}
\address{Department of Mathematics, Tufts University, Medford, Massachusetts 02155, USA}


\begin{abstract}
The Bhatnagar-Gross-Krook (BGK) Boltzmann equation with the Maxwellian-Boltzmann-type equilibrium state leads to the set of thermo-hydrodynamic equations such as the continuity, the Navier-Stokes, and the heat-transfer equations in the scaling limit \cite{1970_Chapman, 1988_Cercignani}.
With its efficient and promising framework handling multi-scale physics, the collision model has been studied with both of theoretical and numerical approaches to apply it for extensive flow conditions such as the flexible choices of the Prandtl number \cite{1966_Holway,1968_Shaknov,2001_Andries,2014_Frapoli,2015_Chen,2018_Atif}.
In this study, using an analytic technique of the kinetic generator \cite{2021_Otomo}, we employ a non-local formulation for the equilibrium state leading to the thermo-hydrodynamic equations with flexible choices of transport coefficients and the equation of state (EOS).
The equilibrium state includes the quasi-equilibrium state intrinsically, being formulated with the non-local macroscopic quantities so that the longer-range interaction is explicitly involved.
According to the new formulation, the consistency with conservation laws, the Boltzmann H-theorem, and the fluctuation-dissipation theorem are examined.
\end{abstract}

\begin{keyword}
\texttt{Bhatnagar-Gross-Krook Boltzmann equation, thermo-hydrodynamic equations, quasi-equilibrium state, Boltzmann-H theorem}
\end{keyword}

\maketitle

\section{Introduction}
\label{introduction}

Dissipative particles' collective motion in the rarefied gas is commonly solved with the Boltzmann equation.
This innovative mathematical representation born from Boltzmann's spirit is continuously evolved until today beyond its original scope.  
The Bhatnagar-Gross-Krook (BGK) collision operator approximates the collision term to circumvent the complex collision operator while maintaining conservation laws and the relaxation process to the equilibrium state \cite{1954_Bhatnagar}.
Employing the Maxwellian distribution as the equilibrium state, one derives the set of thermo-hydrodynamic equations such as the continuity, the Navier-Stokes, and the heat-transfer equation in the scaling limit of the Boltzmann equation.
Based on this fact, studies in these few decades have focused on solving the macroscopic fluid flow using the kinetic-based equation \cite{2001_Succi,2018_Succi}. 
In the derived Navier-Stokes equation and heat-transfer equation, however, the Prandtl number is unity whereas one for a monoatomic gas is approximately $2/3$ \cite{1970_Chapman, 1988_Cercignani}.  
In addition, their transport coefficients, such as viscosity and thermal conductivity, linearly depend on the temperature and density.

To address the unity Prandtl-number problem, the ellipsoidal statistical (ES) BGK model \cite{1966_Holway} and the Shaknov BGK model \cite{1968_Shaknov} were proposed. 
In the ES BGK model, the equilibrium state is formulated by replacing the temperature in the Maxwellian distribution by a tensorial form. It allows the variation of viscosity using a model parameter and was proven to guarantee the non-negative distribution and to obey the Boltzmann H-theorem at a certain range of the parameter \cite{2001_Andries}. 
In contrast to the ES BGK model in which the viscosity is a variable for the flexible Prandtl number, in the Shaknov BGK model an additional factor to the Maxwellian distribution is introduced including the heat flux vector and making the thermal conductivity a variable. 
Such models showed promising results in terms of accuracy as well as computational efficiency compared to the Direct Simulation Monte Carlo (DSMC) method \cite{2018_Pfeiffer}. 
Most of these studies, however, focus on the gas flow at a certain high range of Knudsen number, in which the transport coefficients likely depend on the temperature and density strongly. 
As a result, where the Knudsen number is modest, such approaches are possibly inefficient or invalid.
On the other hand, during these few decades, the other promising formulations are presented in the continuum space.
In one of them, a desirable collision operator for the thermohydrodynamic equation was discussed by focusing on the moments of the equilibrium state \cite{1998_Boghosian}. In the other study, using the multiple-relaxation time (MRT) for moments of the non-equilibrium parts projected to the Hermite space, a desirable set of relaxation times was derived for flexible choices of transport coefficients \cite{2008_Shan}.
They successfully show the ways to improve the applicable range of the Prandtl number.

In the discrete space, the kinetic-based numerical method, the lattice Boltzmann method, has been extensively developed to solve the macroscopic thermo-hydrodynamic problems, resulting in several major schools such as the passive-scalar approach \cite{1993_Bartoloni,1997_Shan}, the two distribution approach \cite{1998_He}, and the multispeed approach \cite{1993_Alexander,1997_McNamara}.
In contrast to the first two approaches, in the multispeed approach the Boltzmann equation is solved for a distribution function that carries density, momentum, and temperature. 
In spite of its physical soundness, however, the multispeed approach requires a number of discrete particles' velocity, which easily causes instability and complexity. 
Some of the recent versions significantly enhanced the robustness by adding the thermal force to the hydrodynamic equation \cite{2009_Sbragaglia}, regularizing the collision operator with the Hermite polynomial \cite{2014_Chen, 2020_Shan}, or using the H-theorem together with the quasi-equilibrium model \cite{2014_Frapoli} and the body-centered-cubic lattices \cite{2018_Atif}.
For example, in some of these studies \cite{2014_Frapoli,2018_Atif}, the equilibrium state was derived from the minimization of the H-function in the discrete space.
This methodology is known to have benefits in terms of robustness and therefore extensively studied with various forms of the H-function such as one based on the Tsallis entropy  \cite{1988_Tsallis, 1999_Boghosian}.
In addition to the inevitable complexity from a number of particles' speeds, however, they sometimes suffer from the limitations such as temperature variation, automatically adjusted viscosity for obeying the H-theorem, or the constant Prandtl number.
Even if they can be addressed using numerical treatments, it is not straightforward to figure out the correspondent formulation in the continuum space from the models in the discrete space.

In another direction, the BGK Boltzmann equation is studied as a kinetic generator for a broad class of partial differential equations in the discrete space as well as the continuum space \cite{2017_Otomo,2018_Otomo,2019_Otomo,2021_Otomo}. 
In a study \cite{2021_Otomo}, after properly choosing the equilibrium state in the continuum space, the BGK Boltzmann equation was solved with the finite-element method and successfully showed consistent solutions with the Burgers', Korteweg-de Vries, and Kuramoto-Sivashinsky equations. 
The keys to this success are to introduce the scale-dependent equilibrium state and to elucidate the detailed high-order structures  in the Chapman-Enskog expansion with the aids of computational tests.

In this study, using the technique in \cite{2021_Otomo} and considering the BGK Boltzmann equation as a kinetic-based representation for the distribution function of the ordinary fluids, we discuss a mathematical form of the equilibrium state in the continuum space for solving the macroscopic thermo-hydrodynamic equations whose transport coefficients and the EOS can be flexibly set without the density and temperature dependence.
Differing from the previous studies in which the collision form was derived for the flexible choices of the Prandtl number \cite{1966_Holway,1968_Shaknov,1998_Boghosian,2008_Shan,2015_Chen} or was derived from the H-theorem \cite{2014_Frapoli,2018_Atif}, the equilibrium state in this study is formulated so that the non-local effects are explicitly included in the equilibrium state with the context of the multi-relaxation processes.
According to the model in this study, we discuss the consistency with conservation laws, the Boltzmann H-theorem and the fluctuation-dissipation theorem.
This paper is organized as follows.
In Sec.~\ref{Maxwellian-introduction}, a mathematical form of the equilibrium state in the BGK Boltzmann equation is presented.
In Sec.~\ref{Derivation_of_Navier_Stokes_and_heat_transfer_equations}, the continuity, the Navier-Stokes, and the heat-transfer equations are derived with the Chapman-Enskog expansion using the non-local equilibrium state.
In Sec.~\ref{roles_of_correction_terms_and_conservation_laws_H_theorem}, the consistency with conservation laws, the the Boltzmann H-theorem, and the fluctuation-dissipation theorem are discussed.
In Sec.~\ref{discussion}, features of the present model, the relation with previous studies, and its possible application are discussed.

\section{A non-local quasi-equilibrium state for the thermo-hydrodynamic equations}
\label{Maxwellian-introduction}

The Boltzmann equation for a particle distribution function $f$ with the BGK collision operator is written as;
\begin{equation}
\label{BGK-Boltzmann}
\frac{\partial f}{\partial t} + c v_j \frac{\partial f}{\partial x_j} = - \frac{f - f^{eq}}{\tau},
\end{equation}
where $\tau$ is the relaxation time and $v_j$ is the particle velocity.
Through this paper, all repeated indices imply summation over them.
Here, all quantities are non-dimensionalized using the characteristic length $l_0$ and time $t_0$.
For example, in the rarefied gas, $l_0$ is chosen as the mean-free path and $t_0$ is chosen so that the characteristic particle velocity is equivalent to $l_0/t_0$.
If the case of ordinary fluids is considered, it is not necessary to choose the characteristic propagation speed of $f$ as the characteristic particle velocity.
Accordingly, the scaling factor $c$ is explicitly written in the second term of the left-hand side. 
Because the propagation speed of $f$ cannot exceed the particle velocity, $c$ should be less than 1.

The Maxwellian-Boltzmann distribution in $D$ spatial dimensions for density $\rho$ and temperature $T$ is,
\begin{equation}
\label{gaussian}
f^{eq} \left( \rho, v, T \right)= \frac{\rho}{ \left(2 \pi T  \right)^{D/2}} \exp \left( - \frac{\left| v-u \right|^2}{2  T} \right),
\end{equation}
where the unit of $T$ is properly chosen so that particle's mass $m$ and the Boltzmann constant $k_B$ are set as unity.
The equilibrium state, Eq.~(\ref{gaussian}), is the solution of the Boltzmann equation with the original complex collision operator, whose application is usually justified for the rarefied gas.  
In previous studies \cite{1970_Chapman, 1988_Cercignani}, the derived Navier-Stokes and the heat-transfer equations using Eq.~(\ref{gaussian}) have transport coefficients linearly depending on density and temperature.

The collision term, the right-hand side in Eq.~(\ref{BGK-Boltzmann}), indicates that the distribution function $f$ changes into an equilibrium state $f^{eq}$ with a single relaxation time $\tau$. 
Here we try to consider multiple-relaxation processes, for example, a collision process in which $f$ changes into a equilibrium state $f^{eq}$ via a equilibrium state $g^{(eq,1)}$ in a different scale. 
The relaxation process is assumed to be characterized by two relaxation times $\tau_A$ and $\tau_B$ where $\tau_A < \tau_B$. 
Then the collision term may be written as,
\begin{equation}
\label{multiple-relaxation}
-\frac{f - g^{(eq,1)}}{\tau_{A}}- \frac{g^{(eq,1)}-f^{eq}}{\tau_{B}}.
\end{equation}
Similar collision processes have been already discussed in many previous studies, for example \cite{2007_Ansumali}, with a different context in the lattice Boltzmann theory. 
After arrangements, we obtain,
\begin{equation}
 \mbox{Eq.~(\ref{multiple-relaxation}) }=  -\frac{f - g^{(eq,1)} \left( 1- \frac{\tau_{A}}{\tau_{B}} \right)  -   f^{eq} \frac{\tau_{A}}{\tau_{B}}  }{\tau_{A}} =-\frac{f - f^{(eq,new)}}{\tau_{A}},
\end{equation}
where $f^{(eq,new)} = g^{(eq,1)} \left( 1- \frac{\tau_{A}}{\tau_{B}} \right)  +   f^{eq} \frac{\tau_{A}}{\tau_{B}}$.
As a result, the new effective equilibrium state, $f^{(eq,new)}$ called as the quasi-equilibrium state or the non-local equilibrium state in this paper, includes two equilibrium states and has dependence on the relaxation times of $\tau_{A}$ and $\tau_{B}$.
We may assume that the relaxation process from $g^{(eq,1)}$ to $f^{eq}$ is one from a non-equilibrium state slightly deviating from the local equilibrium state to the local equilibrium state, which is relevant to the dynamics related to the viscosity and thermal conductivity, according to the liner-response theory \cite{Liboff_2003}.
Then $f^{(eq,new)}$ should include the information of viscosity and thermal conductivity.
The example shown here is about two relaxation processes but can be generalized to an arbitrary number of multiple relaxation processes easily.
In this way, the multiple-relaxation processes can be implicitly formulated with the BGK-Boltzmann equation by a proper definition of the equilibrium state.
When each of the relaxation processes occurs in a different scale, the equilibrium state should include the multi-scale information.
For such types of the equilibrium state, the analysis method in  \cite{2021_Otomo,2019_Otomo,2018_Otomo,2017_Otomo}, in which a broad class of hydrodynamic partial differential equations was generated from the BGK-Boltzmann equation, is useful and will be applied in Sec.~\ref{Derivation_of_Navier_Stokes_and_heat_transfer_equations}.

In this study,  a following equilibrium state is employed to derive the continuity, the Navier-Stokes, and the heat-transfer equations with flexible options of transport coefficients and the EOS,
\begin{equation}
\label{new_eq_def}
\tilde{f}^{eq} = \frac{\rho}{ \left( 2 \pi  \tilde{T}\right)^{D/2}} \exp \left( - \frac{\left| v-\tilde{u} \right|^2}{2 \tilde{T}} \right) + \delta f^{eq},
\end{equation}
where $\tilde{u}$, $\tilde{T}$, and $\delta f^{eq}$ are renormalized local velocity, temperature, and a correction part from the Gaussian distribution defined as followings,
\begin{align}
\label{u_renormalization}
\rho \tilde{u}_i  = \rho u_{i} +  \beta c \tau_1  \frac{\partial E_p}{\partial x_i}, \\
\label{T_renormalization}
\rho \tilde{T}  = \rho T  + \beta c \tau_2 \left\{ E_p  \frac{\partial u_j}{\partial x_j}
- \frac{c \tau_1}{2 \rho} \left|  \frac{\partial E_p}{\partial x_i}  \right|^2 \right\} , \\
\label{delta_feq}
\delta f^{eq} = - \beta f^{eq} \frac{\tilde{\mu} }{\rho T^2 } \left\{ \left(v_l -u_l \right) \left(v_k -u_k \right)  - \frac{\left( v-u \right)^2}{D} \delta_{l,k} \right\} \frac{\partial u_l}{\partial x_k} \nonumber \\
- \beta f^{eq} \frac{ \tilde{\kappa} }{4 \rho T^3 } \left\{ \left( v -u \right)^2 - (D+2)  T \right\} \left( v_k - u_k \right) \frac{\partial T}{\partial x_k},
 \end{align}
 where $E_p=\rho T - P$, $\beta$ is a factor for additional terms, $\mu$ is the dynamic viscosity, $\kappa$ is the thermal conductivity, 
$\tilde{\mu}=\mu - c \tau P$, and $\tilde{\kappa}=\kappa - 2 c \tau P$. 
Here the unit of temperature $T$ is properly chosen so that $m$, $k_B$, and  the thermal capacity at constant volume $C_v$ are set as unity.
By assigning the proper formula to the pressure, $P$, the EOS can be implemented.
The relaxation times $\tau_1$ and $\tau_2$ are introduced in Sec.~\ref{Derivation_of_Navier_Stokes_and_heat_transfer_equations} with the MRT scheme but through the derivation we find that $\tau_1=\tau_2=\tau$ is desirable to be assumed. 
Here, $\sigma$ is the ordinary stress tensor and $\Pi$ is a stress tensor defined as the following,
 \begin{align}
 \label{sigma_def}
\sigma_{i,j}:= \mu \tilde{\sigma}_{i,j} \left( u \right) + \lambda_b \frac{2}{D} \frac{\partial u_l}{\partial x_l} \delta_{i,j}, \\
 \label{tilde_sigma_def}
 \tilde{\sigma}_{i,j} :=\frac{\partial u_j}{\partial x_i}+\frac{\partial u_i}{\partial x_j} - \frac{2}{D} \frac{\partial u_l}{\partial x_l} \delta_{i,j}, \\
\Pi_{i,j} :=\frac{2}{D}  \rho T \tau_1 \tilde{\sigma}_{i,j},
\end{align}
where $\mu$ is the dynamic viscosity and $\lambda_b$ is the bulk viscosity.
Obviously, Eq.~(\ref{new_eq_def})  satisfies the Galilean invariance. 
In addition, with the $\epsilon$ ordering in Sec.~\ref{Derivation_of_Navier_Stokes_and_heat_transfer_equations}, it guarantees the non-negativity of the distribution.
For example, in Eq.~(\ref{T_renormalization}), the first term is positive in the leading order of $\epsilon$ and should be larger than the absolute value of the rest of the terms.

\section{Derivation of the continuity, the Navier-Stokes and the heat-transfer equations}
\label{Derivation_of_Navier_Stokes_and_heat_transfer_equations}

In this section, using the non-local equilibrium state in Eq.~(\ref{new_eq_def}), the continuity, the Navier-Stokes, and the heat transfer equations are derived from the BGK-Boltzmann equation.
For the sake of simplicity, the case in two spatial dimensions is considered. 
To take various relaxation forms into account, the collision term is written with the MRTs $\tau_{m,n}$,
\begin{eqnarray}
\label{MRT-Boltzmann}
\frac{\partial f}{\partial t} + c v_j \frac{\partial f}{\partial x_j} = - \sum_{m \ge 0, n >0} \frac{\mathcal{M}_{m,n} \mathcal{H}_{m,n} e^{- v^2}/ \mathcal{C}_{n,m}}{\tau_{m,n}},
\end{eqnarray}
where
\begin{eqnarray}
\mathcal{M}_{m,n}:=\int \int \left( f - f^{eq}\right)  \mathcal{H}_{m,n} dv_x d v_y,  \\
\mathcal{C}_{m,n}:=\pi 2^{m+n} m! n!,
\end{eqnarray}
and $\mathcal{H}$ is the Hermite polynomial.
Integrating Eq.~(\ref{MRT-Boltzmann}) for $v_x$ and $v_y$ after multiplying $ \mathcal{H}_{l,k}$, one obtains,
\begin{eqnarray}
\label{MRT-Boltzmann-d}
\frac{\partial d_{l,k}}{\partial t} + c \frac{\partial < \mathcal{H}_{l,k}| v_j f>}{\partial x_j}= - \frac{d_{l,k}-d^{eq}_{l,k}}{\tau_{l,k}},
\end{eqnarray}
where
\begin{eqnarray}
\label{d_form}
d_{l,k}= \int \int f \mathcal{H}_{l,k} dv_x dv_y,  \\
d^{eq}_{l,k}= \int \int f^{eq} \mathcal{H}_{l,k} dv_x dv_y.  \label{deq_def}
\end{eqnarray}
Here, the orthogonality of the Hermite polynomial, $\int \int \mathcal{H}_{l,k} \mathcal{H}_{m,n} e^{- v^2} dv_x dv_y = \mathcal{C}_{m,n} \delta_{l,m} \delta_{k,n}$, is used.

Next, the Chapman-Enskog expansion is applied to Eq.~(\ref{MRT-Boltzmann-d}). 
Time and space are ordered by a small dimensionless scaling parameter $\epsilon$.  
The time derivative is ordered as $\partial_t = \sum_{k=1}^\infty \epsilon^k \partial_{t_k}$  where $t_k$ denotes the $k$-th order time scale.  The spatial derivative is ordered as $\partial_x= \epsilon \partial_{x_1}$.  
In the Chapman-Enskog method, gradients of the conserved quantities are assumed to be small.
In addition to the expansion for $f$,
\begin{eqnarray}
f  = \sum_{l=0}^{\infty} \epsilon^l f^{(l)},
\end{eqnarray}
the equilibrium state $f^{eq}$ itself is assumed to depend on the scaling parameter $\epsilon$ as \cite{2021_Otomo},
\begin{eqnarray}
f^{eq}  = \sum_{l=0}^{\infty} \epsilon^l f^{(eq,l)}.
\end{eqnarray}
Accordingly, $d_{n,m}$ and $d^{eq}_{n,m}$ in Eq.~(\ref{d_form}) and Eq.~(\ref{deq_def})  are expanded as,
\begin{eqnarray}
d_{n,m}= \sum_{l=0}^{\infty} \epsilon^l d_{n,m}^{(l)},  \\
d^{eq}_{n,m}= \sum_{l=0}^{\infty} \epsilon^l d_{n,m}^{(eq,l)},  
\end{eqnarray}
where $d_{n,m}^{(l)}$ and $d_{n,m}^{(eq,l)}$ are composed of $f^{(l)}$ and $f^{(eq,l)}$, respectively.
At the $\epsilon^0$ order, Eq.~(\ref{MRT-Boltzmann-d}) can be written as,
\begin{eqnarray}
\label{zero-th-order}
d_{l,k}^{(0)} = d_{l,k}^{(eq,0)}.
\end{eqnarray}
At the $\epsilon$ order, Eq.~(\ref{MRT-Boltzmann-d}) can be written as,
\begin{eqnarray}
\label{first-order}
\frac{\partial d^{(0)}_{l,k}}{\partial t_{1}} + c \frac{\partial < \mathcal{H}_{l,k}| v_j f^{(0)}>}{\partial x_{1,j}}= - \frac{d^{(1)}_{l,k}-d^{(eq,1)}_{l,k}}{\tau_{l+k}}.
\end{eqnarray}
Henceforth, the relaxation times are assumed to be identical if the sum of their two indices is identical. 
With this assumption, the relaxation time is labeled with a single subscript as $\tau_{l+k}$.
Using the recursive relation of the Hermite polynomial,
\begin{eqnarray}
\label{recursive-Hermite}
v_{\gamma} \mathcal{H}_{m,n} = \frac{1}{2} \mathcal{H}_{m+ \delta_{\gamma, x},n + \delta_{\gamma,y}}+ \left( m \delta_{\gamma, x} + n \delta_{\gamma, y} \right) \mathcal{H}_{m- \delta_{\gamma, x},n - \delta_{\gamma,y}} ,
\end{eqnarray}
and notations,
\begin{eqnarray}
\label{D-def}
\mathcal{D}_{i,l,k} [d] :&=& \delta_{i,x} \left(  \frac{d_{l+1,k}}{2} + l  d_{l-1,k} \right) + \delta_{i,y} \left( \frac{d_{l,k+1}}{2} + k d_{l, k-1} \right), \\
\mathcal{D}^{(p)}_{i,l,k} &=& \mathcal{D}_{i,l,k} [d^{(p)}],
\end{eqnarray}
Eq.~(\ref{first-order}) can be written as,
\begin{eqnarray}
\label{first-order_D}
\frac{\partial d^{(0)}_{l,k}}{\partial t_{1}} + c \frac{\partial \mathcal{D}^{(0)}_{j,l,k} }{\partial x_{1,j}}= - \frac{d^{(1)}_{l,k}-d^{(eq,1)}_{l,k}}{\tau_{l+k}}.
\end{eqnarray}
After arrangements, it becomes,
\begin{eqnarray}
\label{first-order_D-arg}
d^{(1)}_{l,k} = d^{(eq,1)}_{l,k} - \tau_{l+k} \left( \frac{\partial d^{(0)}_{l,k}}{\partial t_{1}}  + c \frac{\partial \mathcal{D}^{(0)}_{j,l,k} }{\partial x_{1,j}} \right).
\end{eqnarray}

At the $\epsilon^2$ order, Eq.~(\ref{MRT-Boltzmann-d}) can be written as,
\begin{eqnarray}
\label{sec-order}
\frac{\partial d^{(0)}_{l,k}}{\partial t_{2}} + \frac{\partial d^{(1)}_{l,k}}{\partial t_{1}} + c  \frac{\partial \mathcal{D}^{(1)}_{j,l,k}}{\partial x_{1,j}}
= - \frac{d^{(2)}_{l,k}-d^{(eq,2)}_{l,k}}{\tau_{l+k}}.
\end{eqnarray}
Using Eq.~(\ref{first-order_D-arg}), $\mathcal{D}^{(1)}_{i,l,k}$ can be written as,
\begin{eqnarray}
\label{D_eq}
\mathcal{D}^{(1)}_{i,l,k} = \mathcal{D}^{(eq,1)}_{i,l,k} - \left(   \frac{\partial  \tilde{\mathcal{D}}_{i,l,k} [d^{(0)}]}{\partial t_{1}}  + c \frac{\partial \tilde{\mathcal{D}}_{i,l,k} [\mathcal{D}^{(0)}_{j,l,k}] }{\partial x_{1,j}}  \right),
\end{eqnarray}
where
\begin{eqnarray}
\label{D-def-new}
\mathcal{D}^{(eq,1)}_{i,l,k} &=& \mathcal{D}_{i,l,k} [d^{(eq,1)}], \\
\label{D-tilde-def}
\tilde{\mathcal{D}}_{i,l,k} [d] &=& \delta_{i,x} \left(  \frac{d_{l+1,k}}{2} \tau_{l+1+k} + l  d_{l-1,k} \tau_{l-1+k} \right) + \delta_{i,y} \left( \frac{d_{l,k+1}}{2} \tau_{l+k+1} + k d_{l, k-1} \tau_{l+k-1} \right).
\end{eqnarray}
Therefore,
\begin{eqnarray}
 \tilde{\mathcal{D}}_{i,l,k} [\mathcal{D}_{j,l,k}] = \delta_{i,x} \left(  \frac{\mathcal{D}_{j,l+1,k}}{2} \tau_{l+1+k} + l  \mathcal{D}_{j,l-1,k} \tau_{l-1+k} \right) + \delta_{i,y} \left( \frac{\mathcal{D}_{j,l,k+1}}{2} \tau_{l+k+1} + k \mathcal{D}_{j,l, k-1} \tau_{l+k-1} \right).
\end{eqnarray}
As a result, substituting Eq.~(\ref{D_eq})  to Eq.~(\ref{sec-order}), we obtain,
\begin{eqnarray}
\label{sec-order-last}
\frac{\partial d^{(0)}_{l,k}}{\partial t_{2}} + \frac{\partial d^{(1)}_{l,k}}{\partial t_{1}} 
+ c  \frac{\partial \mathcal{D}^{(eq,1)}_{j,l,k}}{\partial x_{1,j}}
-c \left( \frac{\partial^2 \tilde{\mathcal{D}}_{j,l,k} [d^{(0)}]  }{\partial t_1 \partial x_{1,j}}   
+ c  \frac{\partial^2 \tilde{\mathcal{D}}_{j,l,k} [\mathcal{D}^{(0)}_{i,l,k}]  }{\partial x_{1,i} \partial x_{1,j}}  \right)
= - \frac{d^{(2)}_{l,k}-d^{(eq,2)}_{l,k}}{\tau_{l+k}}.
\end{eqnarray}

Next, Eq.~(\ref{first-order_D}) and Eq.~(\ref{sec-order-last}) at several choices of $l$ and $k$ are considered.
For convenience,  in \ref{deriv_table} symbolical forms of $d^{(eq)}_{l,k}$ are derived and summarized in Table~\ref{tab:moment_new_eq}.
The second, third, and fourth columns show the derived results at each $\epsilon$ order. 

Through the following derivations, whose details are presented in \ref{deriv_cont}, \ref{deriv_NS}, and \ref{deriv_heat}, we find out that  $\tau_0 = \tau_1 = \tau_2 = \tau_3 = \tau$ is a desirable condition for deriving the continuity, the Navier-Stokes, and the heat-transfer equation. 
Therefore the MRT scheme is not effective in this study for deriving the thermo-hydrodynamic equations. 
When the high-order dynamics is explored in the future, however, it may help.


Where $\left\{ l, k \right\}= \left\{ 0,0 \right\}$,  Eq.~(\ref{first-order_D}) and  Eq.~(\ref{sec-order-last}) can be written as,
\begin{eqnarray}
\label{continuity_eq_LO}
\frac{\partial \rho}{\partial t_1} + c  \frac{\partial \rho u_j}{\partial x_{1,j}} =0, \\
\label{continuity_eq_HO}
\frac{\partial \rho}{\partial t_2} = 0,
\end{eqnarray}
respectively.
Assembling them, we obtain the continuity equation,
\begin{eqnarray}
\label{derived_continuity_eq}
\frac{\partial \rho}{\partial t} + c  \frac{\partial \rho u_j}{\partial x_j} =0,
\end{eqnarray}
with the second $\epsilon$ order accuracy.

\

Where $\left\{ l, k \right\}= \left\{ 1,0 \right\}$,  Eq.~(\ref{first-order_D}) is written as,
\begin{eqnarray}
\label{NS_eq_LO_temp}
\frac{\partial \rho u_x}{\partial t_1} + c \frac{\partial \rho u_x u_j}{\partial x_{1, j}}  = -c \frac{\partial}{\partial x_1} \left(  \rho T - \beta \left(  \rho T -P \right) \right).
\end{eqnarray}
When $\beta =1$, the right hand side becomes $-c \frac{\partial P}{\partial x_1} $. 
On the other hand, in the case of ideal monoatomic gas, namely $P=\rho  T$,  the right hand side also becomes $-c \frac{\partial P}{\partial x_1}$ when $\beta=0$. 
As a result, for both cases, Eq.~(\ref{NS_eq_LO_temp}) can be written as,
\begin{eqnarray}
\label{NS_eq_LO}
\frac{\partial \rho u_x}{\partial t_1} + c \frac{\partial \rho u_x u_j}{\partial x_{1,j}}  = -c \frac{\partial P}{\partial x_1}.
\end{eqnarray}

Where $\left\{ l, k \right\}= \left\{ 1,0 \right\}$, Eq.~(\ref{sec-order-last}) is written as,
\begin{eqnarray}
\label{NS_eq_HO}
\frac{\partial \rho u_x}{\partial t_2} = c \frac{\partial  \sigma_{1,x,j} }{\partial x_{1,j}},
\end{eqnarray}
where $\beta=1$ and,
\begin{eqnarray}
\label{NS_eq_HO2}
\frac{\partial \rho u_x}{\partial t_2} = c \frac{\partial  c  \Pi_{1,x,j} }{\partial x_{1,j}},
\end{eqnarray}
in the case of monoatomic ideal gas when $\beta=0$.
Here the first subscript $1$ for $\sigma$ and $\Pi$ denotes the subscript for $x$ and $y$ in all spatial derivatives in  $\sigma$ and $\Pi$.

Assembling Eq.~(\ref{NS_eq_LO}) and Eq.~(\ref{NS_eq_HO}), we obtain the Navier-Stokes equation,
\begin{eqnarray}
\label{NS_eq_derived}
\frac{\partial \rho u_x}{\partial t} + c \frac{\partial \rho u_x u_j}{\partial x_j}  = -c \frac{\partial P}{\partial x} + c   \frac{\partial \sigma_{x,j}}{\partial x_j},
\end{eqnarray}
with the second  $\epsilon$ order accuracy where $\beta=1$. 
As shown in Eq.~(\ref{NS_eq_LO}) and Eq.~(\ref{NS_eq_HO2}), in the case of monoatomic ideal gas at $\beta=0$, Eq.~(\ref{NS_eq_derived}) can be derived as well but its stress tensor has the shear viscosity of $c \tau P$ and zero bulk viscosity , which are consistent with previous studies \cite{1988_Cercignani, 1970_Chapman}.

\

Taking the sum of Eq.~(\ref{first-order_D}) at  $\left\{ l, k \right\}= \left\{ 2,0 \right\}$  and  Eq.~(\ref{first-order_D}) at $\left\{ l, k \right\}= \left\{ 0,2 \right\}$,  we obtain,
\begin{eqnarray}
\label{thermal_eq_LO}
\frac{\partial  \rho T}{\partial t_1} + c \frac{\partial u_j  \rho T}{\partial x_{1,j}}= -c P \frac{\partial u_j}{\partial x_{1,j}},
\end{eqnarray}
when $\beta=1$ and in the case of monoatomic ideal gas when $\beta=0$.

Taking the sum of Eq.~(\ref{sec-order-last}) at  $\left\{ l, k \right\}= \left\{ 2,0 \right\}$  and Eq.~(\ref{sec-order-last}) at $\left\{ l, k \right\}= \left\{ 0,2 \right\}$,  we obtain,
\begin{eqnarray}
\label{thermal_eq_HO1}
\frac{\partial  \rho T}{\partial t_2} = c \left(  u_k \frac{\partial \sigma_{1,j,k}}{\partial x_{1,j}}  + \frac{\partial }{\partial x_{1,j}} \left(  \kappa \frac{\partial T}{\partial x_{1,j}} \right) \right),
\end{eqnarray}
when $\beta=1$,
\begin{eqnarray}
\label{thermal_eq_HO2}
\frac{\partial  \rho T}{\partial t_2} = c  \left(   -  \frac{\partial c   u_k \Pi_{1,j,k} }{\partial x_{1,j}}  + 2 \frac{\partial }{\partial x_{1,j}} \left( c \tau P   \frac{\partial T}{\partial x_{1,j}} \right) \right),
\end{eqnarray}
in the case of monoatomic ideal gas when $\beta=0$.
Assembling Eq.~(\ref{thermal_eq_LO}) and Eq.~(\ref{thermal_eq_HO1}),  we obtain the heat-transfer equation,
\begin{eqnarray}
\label{thermal_eq}
\frac{\partial  \rho T}{\partial t} + c  \frac{\partial u_j  \rho T}{\partial x_j}= -c P  \frac{\partial u_j}{\partial x_j} + c \left(  u_k \frac{\partial \sigma_{j,k}}{\partial x_j}  + \frac{\partial }{\partial x_j} \left(  \kappa \frac{\partial T}{\partial x_j} \right) \right),
\end{eqnarray}
with the second $\epsilon$ order accuracy when $\beta=1$. 
As shown in Eq.~(\ref{thermal_eq_LO}) and Eq.~(\ref{thermal_eq_HO2}), in the case of monoatomic ideal gas when $\beta=0$, Eq.~(\ref{thermal_eq}) can be derived as well but
 the thermal conductivity $\kappa$ is $2 c \tau P$ and the stress tensor has shear viscosity of $c \tau P$ and zero bulk viscosity, which are consistent with previous studies \cite{1988_Cercignani, 1970_Chapman}.


\section{The conservation laws, the Boltzmann H-theorem, and the fluctuation-dissipation theorem}
\label{roles_of_correction_terms_and_conservation_laws_H_theorem}

Conservation laws for mass, momentum, and energy are discussed according to the non-local equilibrium state, Eq.~(\ref{new_eq_def}). 
They can be checked by taking moments of the collision term, specifically the difference between $f$ and $\tilde{f}^{eq}$.
The zeroth moment of $f$ and $\tilde{f}^{eq}$ are obviously $d_{0,0}=d^{eq}_{0,0}=\rho$ as shown in \ref{deriv_table}.
The first-rank moment of  $f$ and $\tilde{f}^{eq}$ are  $d_{1,0}= 2 \rho u_x$ and $d^{eq}_{1,0}= 2 \rho u_x + 2 \beta c \tau_1 \frac{\partial E_p}{\partial x}$ as shown in \ref{deriv_table}. 
The second term of $d^{eq}_{1,0}$ comes from the renormalized velocity in Eq.~(\ref{u_renormalization}),
\begin{align}
\rho \tilde{u}_i = \rho u_{i} + \underbrace{\beta c \tau_1 \frac{\partial E_p}{\partial x_i}}_{uterm1}.  \nonumber
\end{align}
The correction term, $''uterm1''$  contributes to the pressure term in the Navier-Stokes equation as shown in \ref{deriv_NS}, replacing the pressure force of the monoatomic ideal gas by the pressure force from a specified EOS. 
In addition, $''uterm1''$ helps for achieving the second-order accuracy in the continuity equation.
Because $''uterm1''$ disappears when it is integrated over the space with the zero boundaries' contributions, the conservation of momentum is guaranteed for any $\beta$.
The second-rank moment of  $f$ and $\tilde{f}^{eq}$ are  $d_{2,0} +d_{0,2} = 8E -4 \rho $ and $d^{eq}_{2,0} +d^{eq}_{0,2} = 8 \tilde{E}-4 \rho$ where $\tilde{E}=  \rho \tilde{T} + \frac{1}{2} \rho \tilde{u}^2$ as shown in \ref{deriv_table}. 
Here, the renormalized temperature is defined in Eq.~(\ref{T_renormalization}), 
\begin{align}
\rho \tilde{T}  = \rho T  + \beta c \tau_2 \left\{ \underbrace{E_p  \frac{\partial u_j}{\partial x_j}}_{Tterm1}
- \underbrace{ \frac{c \tau_1}{2 \rho} \left|  \frac{\partial E_p}{\partial x_i}  \right|^2}_{Tterm2} \right\}.  \nonumber
 \end{align}
The first correction term, $''Tterm1''$, replaces the compression term in Eq.~(\ref{thermal_eq}) from the ideal monoatomic gas by the compression term from the ordinary fluids with the same motivation as $'''uterm1''$.
The last correction term, $''Tterm2''$, removes the unnecessary non-linear term originated from $\rho \tilde{u}^2$ for the  energy conservation. 
As shown in \ref{deriv_table}, using  Eq.~(\ref{u_renormalization}) and Eq.~(\ref{T_renormalization}),  $d^{eq}_{2,0} +d^{eq}_{0,2} = 8  \rho T + 4 \rho u^2 - 4 \rho +8 c \tau \frac{\partial E_p u_j}{\partial x_j} $. 
The last term disappears after the integration over the space. 
 As a result, the spatial integration for $d_{2,0} + d_{0,2}$ is equivalent to one for $d^{eq}_{2,0} +  d^{eq}_{0,2}$. 
 It guarantees the conservation of the energy $E$.

The consistency with the Boltzmann H-theorem brings an advantage for numerical stability \cite{2014_Frapoli,2018_Atif, 1999_Karlin, 2001_Boghosian, 2018_Wilson}.
It is also well known that the BGK Boltzmann equation with the Maxwellian distribution for the equilibrium state satisfies the H-theorem. 
Namely, the following functional $\mathcal{H}$ monotonically decreases,
\begin{eqnarray}
\label{H_fnc}
\mathcal{H} = \int \int f \ln  f  dv dx,
\end{eqnarray}
and exhibits its minimum where $f=f^{eq}$.
For the non-local equilibrium state, Eq.~(\ref{new_eq_def}), using the conservation of mass, momentum, and energy, the monotonic behavior of $\mathcal{H}$ toward $\tilde{f}^{eq}$ can be shown if the deviation of the pressure formulation from the ideal-gas EOS, $E_p$, is sufficiently small, $E_p \le \mathcal{O} \left( \epsilon \cdot \max \left\{ \tilde{\mu}, \tilde{\kappa} \right\}  / c \tau \right)$.
To show the monotonic behavior of $\mathcal{H}$, after multiplying $\ln f +1$ to the BGK Boltzmann equation, Eq.~(\ref{BGK-Boltzmann}), and taking integration for $v$ and $x$, one can derive,
\begin{eqnarray}
\frac{d \mathcal{H}}{d t} = - \frac{1}{\tau} \int \int \left( f -\tilde{f}^{eq} \right) \left( \ln f - \ln \tilde{f}^{eq} \right) dv dx -  \frac{1}{\tau} \int \int \left( f -\tilde{f}^{eq} \right) \ln \tilde{f}^{eq} dv dx.
\end{eqnarray}
Due to the monotonicity of the logarithm, the first kernel in the right-hand side is negative. 
If the second kernel is expanded by $\epsilon$, its leading order is,
 \begin{eqnarray}
 - \int \int \left( f -\tilde{f}^{eq} \right) \ln \tilde{f}^{eq} dv dx \approx  -  \epsilon^2  \int \int    \left( f^{(1)} -\tilde{f}^{(eq,1)}\right)  \frac{\tilde{f}^{(eq,1)}}{\tilde{f}^{(eq,0)}}  dv dx \le 0,
\end{eqnarray}
whose derivation is shown in \ref{appendix_deriv_Hth} in the case of $\tilde{\mu} \ge 0$ and $\tilde{\kappa} \ge 0$ which can be achieved by adjusting $c$ and $\tau$ if necessary.
As a result, $\frac{d \mathcal{H}}{d t}  \le 0$.
If we consider the following functional $\mathcal{L}$ for minimizing $\mathcal{H}$ under the mass, momentum, and energy conservation,
\begin{eqnarray}
\mathcal{L}= \int \int \left( f \ln f + l_0 f + l_1 f v + l_2 f v^2 \right) dv dx,
\end{eqnarray}
where $l_0$, $l_1$, and $l_2$ are the Lagrange multipliers, then we find that $f$ satisfying  $\frac{\delta \mathcal{L}}{\delta f}=0$ takes the Gaussian form.  
Therefore, the equilibium state of Eq.~(\ref{new_eq_def}) can meet this criteria with a proper choice of $l_0$, $l_1$, and $l_2$ if  $\delta f^{eq}$ is ignored.
Because $\delta f^{eq}$ contributes to the dissipation where $\tilde{\mu} \ge 0$ and $\tilde{\kappa} \ge 0$ and disappears as time goes, the functional $\mathcal{H}(f)$ likely monotonically decreases toward  $\mathcal{H}( f^{eq})$ via $\mathcal{H}( \tilde{f^{eq}})$.
 
Next, the consistency with the fluctuation-dissipation theorem is examined.
In this paragraph, we assume $E_p \le \mathcal{O} \left( \epsilon \right)$ and choose an inertial frame in which the local fluid velocity $u \vert_{x}$ is zero.
Using the x-y component of a microscopic pressure tensor $P_{xy}$ and microscopic heat flux $\mathcal{S}_i$, the transport coefficients are formulated in two spatial dimensions as, 
\begin{align}
\label{FDT_mu}
\mu= \frac{c}{ T} \int ^{\infty}_{0} \langle P_{xy} \left( t \right)  P_{xy} \left( 0 \right) \rangle dt, \\
\label{FDT_kappa}
\kappa = \frac{c}{2  T^2} \int ^{\infty}_{0} \langle \mathcal{S}_i (t) \mathcal{S}_i (0)  \rangle dt.
\end{align}
Here the angle bracket denotes the time-correlation function.
Assuming $f$ is close to $\tilde{f^{eq}}$, one writes the right-hand side in Eq.~(\ref{FDT_mu}) as;
\begin{align}
\label{RHS_FDT}
 \mbox{RHS of Eq.~(\ref{FDT_mu})}=  \frac{c}{ T} \int \int \int ^{\infty}_{0} v_x v_y v'_x v'_y   \tilde{f}^{eq}\left( v \right) \mathcal{P} \left( v', t | v, 0 \right)  dt dv dv'.
\end{align}
 Here $\mathcal{P} \left( v', t | v, 0 \right)$ is the conditional probability of finding the particle of velocity $v'$ at time $t$ where the particle of velocity $v$ was found at time 0. It may be written with the functional derivative as,
\begin{align}
\label{conditiona_probab}
\mathcal{P} \left( v', t | v, 0 \right) = \frac{\delta f \left( v', t\right)}{\delta f \left( v, 0\right)}
= \frac{\delta f \vert_{v', t}}{\delta f \vert_{v, 0}} 
+ \int^{t}_{0} \int \frac{\delta f \vert_{v', t}}{\delta \tilde{f}^{eq} \vert_{v'', t'}}  \frac{\delta \tilde{f}^{eq} \vert_{v'', t'}}{\delta f \vert_{v, 0}} dv'' dt'  + \cdots.
\end{align}
The first term in the right-hand side presents the correlation between an initial state and a state after advection.
It should be the exponential decay for time due to the collision during the particle's advection.
The second term in the right-hand side presents the correlation between an initial state and an advected state after a collision. 
Specifically, particles collide after advecting from an initial state with velocity $v$ and then follow the equilibrium state. After that, the equilibrium state advects back to the original position at time $t$ with particle velocity $v'$, which is an opposite number to $v$.
If both of such two terms are explicitly written, they are
\begin{align}
\label{conditiona_probab2}
\mbox{Eq.~(\ref{conditiona_probab})}
= \exp \left( - \frac{t}{\tau} \right) \delta \left( v' - v \right) 
+ \int^{t}_{0} \int \exp \left( - \frac{t-t'}{\tau} \right) \delta \left( v' - v'' \right)   \frac{\delta \tilde{f}^{eq} \vert_{v'', t'}}{\delta X \vert_{t'}}
\frac{ \delta X \vert_{t'}}{\delta f \vert_{ v, 0 }} \delta \left( t'- \tau \right)  dv'' dt' + \cdots.
\end{align}
Here, to simplify the time integration for $t'$, typical time for the collision $\tau$ is picked using the Dirac delta function $\delta \left( t'- \tau \right) $. 
The initial state $f \vert_{ v, 0 }$ contributes to the equilibrium state $ \tilde{f}^{eq} \vert_{v'', t'}$  via the macroscopic physical quantity $X \vert_{t'}$,  $X= \left\{ \rho, v_{\alpha}, T, \nabla_{\alpha} v_{\beta}, \nabla_{\alpha} T \right\}$ where $\alpha$ and $\beta$ are indicies for the spatial direction. 
 Using Eq.~(\ref{conditiona_probab}) and the functional derivative of $X \vert_{t'}$ in terms of $\delta f \vert_{v,0}$,
 \begin{align}
\label{X_Frechet}
 \frac{\delta X \vert_{t'}}{\delta f \vert_{v,0}} 
 = \left\{ \delta\left( t' \right) + \delta \left( v \right) \exp \left( - \frac{t'}{\tau}\right),
  \frac{v_{\alpha} }{\rho} \delta \left( t' \right) ,  \right. \nonumber \\
\left. \frac{v^2 - 2 T}{2 \rho}  \left\{ \delta \left( t' \right) + \delta \left( v \right) \exp \left( -\frac{t'}{\tau}\right) \right\}, 
 -\frac{v_{\beta}}{\rho v_{\alpha} t'} \exp \left( -\frac{t'}{\tau} \right),
 -\frac{1}{  v_{\alpha} \tau} \frac{v^2 - 2 T }{2 \rho  }\exp \left( -\frac{t'}{\tau} \right)
  \right\},
\end{align}
 whose derivation are shown in \ref{appendix_deriv_FDT}, Eq.~(\ref{RHS_FDT}) can be written as,
 \begin{align}
\mbox{Eq.~(\ref{RHS_FDT})} \approx  c \tau \rho T + \frac{\tilde{\mu}}{\rho^2  T^3} \int \int v^2_y \tilde{f}^{eq} (v) v^{\prime 2}_x v^{\prime 2}_y  f^{eq} (v^{\prime}) dv dv^{\prime}
= c \tau \rho T  + \tilde{\mu} \approx \mu,
\end{align}
whose detailed derivations are shown in \ref{appendix_deriv_FDT}. 
Similarly, the right-hand side of Eq.~(\ref{FDT_kappa}) can be written as,
 \begin{align}
\mbox{RHS of Eq.~(\ref{FDT_kappa})}\approx \nonumber \\
\frac{c}{2  T^2} \int \int \int^{\infty}_{0} \left( \frac{v^2_x + v^2_y}{2} - 2  T \right)  \left( \frac{v^{\prime 2}_x + v^{\prime 2}_y}{2} - 2  T \right) \left( v_x v^{\prime}_x + v_y v^{\prime}_y \right) \tilde{f}^{eq} \left( v \right)  \mathcal{P} \left( v', t | v, 0 \right)  dt dv dv^{\prime}, \nonumber \\
\approx 2 c \rho T  \tau + \tilde{\kappa} \approx \kappa
\end{align}
using Eq.~(\ref{conditiona_probab}), Eq.~(\ref{conditiona_probab2}), and Eq.~(\ref{X_Frechet}). Again the detailed derivation is shown in \ref{appendix_deriv_FDT}. 
As a result, the non-local equilibrium state shows consistency with the fluctuation-dissipation theorem where $E_p \le \mathcal{O} \left( \epsilon \right)$.

\section{Discussion}
\label{discussion}

The non-local equilibrium state, Eq.~(\ref{new_eq_def}), successfully leads to the set of thermo-hydrodynamic equations with flexible choices of the transport coefficients and the EOS.
It involves the longer range interaction with the non-local terms in the renormalized fluid velocity and temperature, Eq.~(\ref{u_renormalization}) and Eq.~(\ref{T_renormalization}), and an additional state to the Gaussian distribution, Eq.~(\ref{delta_feq}).
The BGK Boltzmann equation with this equilibrium state holds consistency with the conservative law for mass, momentum, and energy.
In addition, it shows the consistency with the Boltzmann H-theorem where $E_p \le \mathcal{O} \left( \epsilon \cdot \max \left\{ \tilde{\mu}, \tilde{\kappa} \right\}  / c \tau \right)$, and the fluctuation-dissipation theorem where $E_p \le \mathcal{O} \left( \epsilon \right)$. 
Such limitation for $E_p$, the deviation from the ideal-gas EOS, may sound reasonable because the strong particles' interaction, such as realizing the phase separation, can result in the large modification of the EOS.

In contrast to the present model, the collision model in the previous studies \cite{1966_Holway,1968_Shaknov,2014_Frapoli,2018_Atif,2008_Shan,2014_Chen,2020_Shan} involves the non-local information using the second-order and the third-order moments of $f$ and its deviation from $f^{eq}$. 
That is, the following approximate relationship obtained at the $\epsilon$ order of the Chapman-Enskog expansion with an assumption of negligible temporal variation,
\begin{align}
\int \underbrace{ \left( v_{\alpha} v _{\beta} \cdots \right)}_{n} c v_j \frac{\partial f^{(0)}}{\partial x_{1,j}} dv \approx - \int \underbrace{ \left( v_{\alpha} v _{\beta} \cdots \right)}_{n} \left( f^{(1)} - f^{(eq,1)} \right) dv / \tau,
\end{align}
where $n \ge 2$, indicates that the second and the third-order moments of $f$ or its deviation from $f^{eq}$  are proportional to the derivatives of the fluid velocity and temperature in the equilibrium state.
Furthermore, in contrast to the previous studies \cite{2014_Frapoli,2018_Atif,2008_Shan,2014_Chen,2020_Shan}, the present model apparently requires modifying the equilibrium state for changing the viscosity and thermal conductivity instead of modifying the relaxation times. 
As shown in Section.~\ref{Maxwellian-introduction}, however, if the equilibrium state is formulated with the context of the multi-relaxation processes, it is possible to understand that the transport coefficients are set using the relaxation times because they are already involved in the newly defined equilibrium state.
This relaxation scheme seems to be closer to the quasi-equilibrium scheme \cite{2014_Frapoli,2018_Atif,2007_Ansumali} rather than one assigning the different relaxation times for the different set of moments of $f$ or its deviation from $f^{eq}$ \cite{2008_Shan,2014_Chen,2020_Shan}.
Then, the role of $\tau$ appeared in Eq.~(\ref{BGK-Boltzmann}) may be questionable.
If Eq.~(\ref{BGK-Boltzmann}), Eq.~(\ref{u_renormalization}), and Eq.~(\ref{T_renormalization}) are carefully examined,  one can find that $\tau$ and  $c$ can appeared with their products everywhere except for the temporal derivative term in Eq.~(\ref{BGK-Boltzmann}). It indicates that the roles of $c$ and $\tau$ are similar except for effects on the time scale. 
According to the derived macroscopic equations, Eq.~(\ref{derived_continuity_eq}), Eq.~(\ref{NS_eq_derived}), and Eq.~(\ref{thermal_eq}), the reduction of $c$ leads the reduction of time-scale. 
In addition, at higher $\epsilon$ orders than the second-order, higher spatial derivative terms should appear bringing the higher power of $c$ from the advection term in  Eq.~(\ref{BGK-Boltzmann}).
Therefore, the reduction of $c$ leads to the reduction of the truncation error terms in the macroscopic equations.
As a result, the reduction of $\tau$ likely leads to the reduction of the truncation terms' effects without changing the time scale although the choice of its magnitude is limited compared to $c$.
From the viewpoint of the multiple collision processes in Section.~\ref{Maxwellian-introduction}, $\tau$ is relevant to the relaxation from $f$ to $g^{(eq,1)}$, which does not affect the non-equilibrium phenomena related to the viscosity and thermal conductivity but is related to the high $\epsilon$ order physics and stability in the numerical simulation.

The ES BGK and the Shaknov BGK model introduced in Sec.~\ref{introduction} modified the mathematical form of the collision term for flexible choices of the Prandtl number similarly to the present model.  
Some particle methods based on these models have already shown promising results for the gas flow in the nozzle and channels \cite{2006_Burt, 2008_Titov,2018_Pfeiffer}.
They are expected to handle the Knudsen number effects in the slip-flow regime and the transition flow regime.
Due to the density and temperature dependence on the transport coefficients, however, its application for the continuum flow regime is limited.
To establish a kinetic-based approach that can cover a broad range of the Knudsen number, the combination between such models and the present model is possibly promising after adopting the present model to the discrete space for the numerical simulation. 

The theory developed in this study is applicable for analysis of the particle image velocimetry (PIV) and the particle tracking velocimetry (PTV), popular detection tools for the fluid velocity in the experimental fluid dynamics \cite{2007_Raffel,2019_Takikawa, 2021_Vilquin}. 
With recent advancements of the high-speed camera and the discovery of the high refractive materials, the tracer particle's size can be a reduced to few nano-meters, which help to avoid the disturbance to the flow field and to apply for the microfluidic flow \cite{2010_Stuart}. 
Moreover, the pronounced fluctuation due to the Brownian motion provides us more information such as the temperature using the fluctuation-dissipation theorem \cite{2009_Pramod}, however its dependences on the flow and temperature fields are not fully understood yet.
Cercignani formulated the kinetic-based equation with the fluctuations as followings,
\begin{align}
\label{BGK-Ceicigani}
\frac{\partial f}{\partial t} + c v_j \frac{\partial f}{\partial x_j} = - \frac{f - \tilde{f}^{eq}}{\tau} + \frac{\partial^2}{\partial v^2} \left( \bar{D} f\right)+ \frac{\partial}{\partial v_i} \left\{ \left(v_i-u_i \right) \gamma f \right\},
\end{align}
where $\bar{D}$ is a coefficient of the random force's correlation and $\gamma$ is the friction coefficient \cite{1988_Cercignani}.
In the right-hand side, in addition to the BGK collision term, there are the diffusion term and drifting term from the Fokker-Planck (FP) equation.
When the tracer particle's dynamics is much slower than the solvent particles' dynamics due to the differences of particles' mass and size,  the timescale of the BGK collision term $\tau$ is much larger than the timescale of the last two terms, the FP terms. Then the long-range interaction force becomes dominant and Eq.~(\ref{BGK-Ceicigani}) can be approximated by the FP equation.
If these two-time scales are comparable, all terms in Eq.~(\ref{BGK-Ceicigani}) should be considered.
In that case, if  $f= \tilde{f}^{eq}$ is applied to the FP term at the incompressible limit $\partial u_l / \partial x_l =0$, we obtain,
\begin{align}
\label{BGK-Ceicigani2}
 \frac{\partial^2}{\partial v^2} \left( \bar{D}  \tilde{f}^{eq} \right)+ \frac{\partial}{\partial v_i} \left\{ \left(v_i-u_i \right) \gamma \tilde{f}^{eq} \right\}= - 2 \gamma \delta f^{eq},
\end{align}
where $\bar{D}=\gamma T$, $E_p=0$, and temperature variation is negligible. The derivation is shown in \ref{deriv_FPterm}. This non-zero contribution due to the non-locality of the equilibrium state can be integrated by the BGK collision term if the transport coefficients in $\tilde{f}^{eq}$ in Eq.~(\ref{BGK-Ceicigani}) are corrected.
It indicates that the non-local equilibrium state in this study can play a part of the roles of the FP terms where $f$ is close to $\tilde{f}^{eq}$.
As a result, the distribution function should be smoothly relaxated to $f^{eq}$ via around $\tilde{f}^{eq}$ while the fluctuation property is maintained with the non-locality of $\tilde{f}^{eq}$.
Thereby, it is promising to solve or analyze Eq.~(\ref{BGK-Ceicigani}) with the non-local equilibrium state for aid of analysis of experimental observations.

\section{Acknowledgements}

H.O. appreciates Prof. Bruce Boghosian, Prof. Hudong Chen and Prof. Sauro Succi for helpful comments and discussions.


\appendix
\section{Moments of $\tilde{f}^{eq}$ in Eq.~(\ref{new_eq_def})}
\label{deriv_table}

The non-local equilibrium state $\tilde{f}^{eq}$ defined in Eq.~(\ref{new_eq_def}) is projected to the Hermite space in two spatial dimensions.
For convenience, the moments of the Maxwellian-Boltzmann-type distribution $ f^{eq}$ in Eq.~(\ref{gaussian}) are written in followings,
\begin{align}
\label{feq_moment1}
\int \begin{pmatrix} 1 \\ v_x \\ v^2_x \\ v^3_x \\ v^4_x \\ v^5_x \\ v^6_x \end{pmatrix} f^{eq} d v_x d v_y=  
\begin{pmatrix} \rho \\ \rho u_x  \\  \rho T + \rho u^2_x 
\\ 3 \rho T  u _x + \rho u^3_x   
\\ 3 \rho T^2  + 6 \rho u^2_x T  + \rho u^4_x 
\\  15 \rho T^2  u_x + 10 \rho u^3_x T  + \rho u^5_x  
\\  15 \rho T^3   + 45 \rho T^2   u^2_x + 15 \rho T  u^4_x + \rho u^6_x  \end{pmatrix}, \\
\label{feq_moment2}
\int \begin{pmatrix} v_x v_y \\ v_x v^2_y \\ v^2_x v^2_y \\   \end{pmatrix} f^{eq} d v_x d v_y= 
\begin{pmatrix} \rho u_x u_y
\\ \rho u_x \left( T  +  u^2_y \right) 
\\ \rho \left( T  + u^2_x \right) \left( T  + u^2_y \right)  
 \end{pmatrix}.
\end{align}

For $d^{eq}_{0,0}$, the equilibrium state is integrated as,
\begin{align}
d^{eq}_{0,0}= \int \tilde{f}^{eq} d v_x d v_y = \rho.
\end{align}
It is because 
\begin{align}
\int \delta f^{eq} d v_x d v_y =- \beta \frac{\tilde{\mu} }{\rho T^2 } \frac{\partial u_l}{\partial x_k} \int f^{eq} \left\{ \left(v_l -u_l \right) \left(v_k -u_k \right)  - \frac{\left( v-u \right)^2}{2} \delta_{l,k} \right\}  dv_x dv_y \nonumber \\
- \beta  \frac{ \tilde{\kappa} }{4 \rho T^3 }  \frac{\partial T}{\partial x_k}  \int  f^{eq} \left\{ \left( v -u \right)^2 -4  T \right\} \left( v_k - u_k \right)dv_x dv_y
=0,
\end{align}
using Eq.~(\ref{delta_feq}), Eq.~(\ref{feq_moment1}), and Eq.~(\ref{feq_moment2}).

For $d^{eq}_{1,0}$, using Eq.~(\ref{new_eq_def}) and Eq.~(\ref{u_renormalization}), the integration can be done as,
\begin{align}
d^{eq}_{1,0}= 2 \int \int \tilde{f}^{eq} v_x d v_x d v_y = \rho \tilde{u}_x= 2 \rho u_{x} + 2 \beta c \tau_1  \frac{\partial E_p}{\partial x} .
\end{align}
It is because 
\begin{align}
\int \delta f^{eq} v_x d v_x d v_y =- \beta \frac{\tilde{\mu} }{\rho T^2 } \frac{\partial u_l}{\partial x_k} \int v_x f^{eq} \left\{ \left(v_l -u_l \right) \left(v_k -u_k \right)  - \frac{\left( v-u \right)^2}{2} \delta_{l,k} \right\}  dv_x dv_y \nonumber \\
- \beta  \frac{ \tilde{\kappa} }{4 \rho T^3 }  \frac{\partial T}{\partial x_k}  \int  v_x f^{eq} \left\{ \left( v -u \right)^2 -4  T \right\} \left( v_k - u_k \right)dv_x dv_y
=0,
\end{align}
using Eq.~(\ref{delta_feq}), Eq.~(\ref{feq_moment1}), and Eq.~(\ref{feq_moment2}).

For $d^{eq}_{2,0}$, using Eq.~(\ref{u_renormalization}) and Eq.~(\ref{T_renormalization}), we obtain,
\begin{align}
d^{eq}_{2,0}=  \int \int \tilde{f}^{eq}  \left(4  v^2_x -2 \right)  d v_x d v_y = 4 \left(  \rho \tilde{T} + \rho \tilde{u}_x^2 \right)-2 \rho - 2 \beta  \tilde{\mu} \left( \tilde{\sigma}_{x,x} - \tilde{\sigma}_{y,y} \right)  \nonumber \\
= 4 \left(  \rho T + \rho u_x^2 \right)-2 \rho 
+ 4 \beta c \tau_2 \left\{ E_p  \frac{\partial u_j}{\partial x_j}
- \frac{c \tau_1}{2 \rho} \left|  \frac{\partial E_p}{\partial x_i}  \right|^2 \right\} 
\nonumber \\
+ 8 \beta c \tau_1 u_x  \frac{\partial E_p}{\partial x} +  4 \frac{\beta^2 c^2 \tau^2_1}{\rho} \left( \frac{\partial E_p}{\partial x} \right)^2  - 2 \beta  \tilde{\mu} \left( \tilde{\sigma}_{x,x} - \tilde{\sigma}_{y,y} \right)  \nonumber \\
= 4 \left(  \rho T + \rho u_x^2 \right)-2 \rho+ 4 \beta c \tau_2 E_p  \frac{\partial u_j}{\partial x_j} 
+ 8  \beta c \tau_1 u_x  \frac{\partial E_p}{\partial x}  \nonumber \\
- 2 \beta  \tilde{\mu} \left( \tilde{\sigma}_{x,x} - \tilde{\sigma}_{y,y} \right)  
- 2 \frac{ \beta c^2 \tau_2 \tau_1}{\rho} \left| \frac{\partial  E_p}{\partial x_i} \right|^2  
+  4 \frac{\beta^2 c^2 \tau^2_1}{\rho} \left( \frac{\partial E_p}{\partial x}  \right)^2.
\end{align}
It is because 
\begin{align}
\int \delta f^{eq} v^2_x  d v_x d v_y =- \beta \frac{\tilde{\mu} }{\rho T^2 } \frac{\partial u_l}{\partial x_k} \int v^2_x f^{eq} \left\{ \left(v_l -u_l \right) \left(v_k -u_k \right)  - \frac{\left( v-u \right)^2}{2} \delta_{l,k} \right\}  dv_x dv_y \nonumber \\
- \beta  \frac{ \tilde{\kappa} }{4 \rho T^3 }  \frac{\partial T}{\partial x_k}  \int  v^2_x f^{eq} \left\{ \left( v -u \right)^2 -4  T \right\} \left( v_k - u_k \right)dv_x dv_y \nonumber \\
=- \beta \tilde{\mu} \left( \frac{\partial u_x}{\partial x} - \frac{\partial u_y}{\partial y}  \right) 
= - \frac{\beta}{2}  \tilde{\mu} \left( \tilde{\sigma}_{x,x} - \tilde{\sigma}_{y,y} \right)  .
\end{align}
using Eq.~(\ref{delta_feq}), Eq.~(\ref{feq_moment1}), and Eq.~(\ref{feq_moment2}).
Therefore where $\tau_1=\tau_2=\tau$ and $\beta=1$, the sum of $d^{eq}_{2,0}+d^{eq}_{0,2}$ becomes,
 \begin{align}
d^{eq}_{2,0}+d^{eq}_{0,2}= 8  \rho T + 4 \rho u^2 - 4 \rho +8 \beta c \tau \frac{\partial E_p u_j}{\partial x_j}.
\end{align}

For $d^{eq}_{1,1}$, using Eq.~(\ref{u_renormalization}), we obtain,
\begin{align}
d^{eq}_{1,1}= 4 \int \int \tilde{f}^{eq} v_x v_y d v_x d v_y = 4 \rho \tilde{u}_x \tilde{u}_y - 4 \beta  \tilde{\mu} \tilde{\sigma}_{x,y} \nonumber \\
= 4 \rho u_x u_y + 4 \beta c \tau_1 \left\{ u_x  \frac{\partial E_p}{\partial y}   + u_y \frac{\partial E_p}{\partial x}\right\} 
- 4 \beta  \tilde{\mu} \tilde{\sigma}_{x,y}.
\end{align}
It is because 
\begin{align}
\int \delta f^{eq} v_x  v_y  d v_x d v_y 
=- \beta \frac{\tilde{\mu} }{\rho T^2 } \frac{\partial u_l}{\partial x_k} \int v_x v_y f^{eq} \left\{ \left(v_l -u_l \right) \left(v_k -u_k \right)  - \frac{\left( v-u \right)^2}{2} \delta_{l,k} \right\}  dv_x dv_y \nonumber \\
- \beta  \frac{ \tilde{\kappa} }{4 \rho T^3 }  \frac{\partial T}{\partial x_k}  \int  v_x v_y f^{eq} \left\{ \left( v -u \right)^2 -4  T \right\} \left( v_k - u_k \right)dv_x dv_y \nonumber \\
=- \beta   \tilde{\mu} \left( \frac{\partial u_x}{\partial y} + \frac{\partial u_y}{\partial x}  \right)  
= - \beta  \tilde{\mu}  \tilde{\sigma}_{x,y} 
\end{align}
using Eq.~(\ref{delta_feq}), Eq.~(\ref{feq_moment1}), and Eq.~(\ref{feq_moment2}).

For $d^{eq}_{3,0}$, using Eq.~(\ref{u_renormalization}) and Eq.~(\ref{T_renormalization}), we obtain,
\begin{align}
d^{eq}_{3,0}=  \int \int \tilde{f}^{eq} \left( 8 v^3_x - 12 v_x \right) d v_x d v_y = 
8 \rho  \tilde{u}^3_x + 24 \tilde{u}_x  \rho \tilde{T} - 12 \rho \tilde{u}_x -12 \beta  \tilde{\mu} u_x \left( \tilde{\sigma}_{x,x} - \tilde{\sigma}_{y,y} \right) -12 \beta  \tilde{\kappa} \frac{\partial T}{\partial x} \nonumber \\
= 8 \rho  u^3_x + 24 u_x  \rho T - 12 \rho u_x + \left( 24 u^2_x + 24  T -12 \right) \beta c \tau_1 \frac{\partial E_p}{\partial x} + 24 u_x \beta c \tau_2 E_p \frac{\partial u_j}{\partial x_j} \nonumber \\
-12 \beta  \tilde{\mu} u_x \left( \tilde{\sigma}_{x,x} - \tilde{\sigma}_{y,y} \right) -12 \beta  \tilde{\kappa} \frac{\partial T}{\partial x} + H.O.
\end{align}
It is because 
\begin{align}
\int \delta f^{eq} \left( 8 v^3_x - 12 v_x \right)  d v_x d v_y 
=- \beta \frac{\tilde{\mu} }{\rho T^2 } \frac{\partial u_l}{\partial x_k} \int 8 v^3_x f^{eq} \left\{ \left(v_l -u_l \right) \left(v_k -u_k \right)  - \frac{\left( v-u \right)^2}{2} \delta_{l,k} \right\}  dv_x dv_y \nonumber \\
- \beta  \frac{ \tilde{\kappa} }{4 \rho T^3 }  \frac{\partial T}{\partial x_k}  \int  8 v^3_x  f^{eq} \left\{ \left( v -u \right)^2 -4  T \right\} \left( v_k - u_k \right)dv_x dv_y, \nonumber \\
= - 12 \beta  \tilde{\mu} u_x \left( \tilde{\sigma}_{x,x} -  \tilde{\sigma}_{y,y} \right)   - 12 \beta  \tilde{\kappa} \frac{\partial T}{\partial x}.
\end{align}
using Eq.~(\ref{delta_feq}), Eq.~(\ref{feq_moment1}), and Eq.~(\ref{feq_moment2}).

For $d^{eq}_{2,1}$, using Eq.~(\ref{u_renormalization}) and Eq.~(\ref{T_renormalization}), we obtain,
\begin{align}
d^{eq}_{2,1}= \int \int \tilde{f}^{eq} \left( 8  v^2_x v_y - 4 v_y \right) d v_x d v_y  \nonumber \\
=  8 \rho  u^2_x  u_y + 8 u_y  \rho T - 4 \rho u_y
+ 16 u_x u_y \beta c \tau_1 \frac{\partial E_p}{\partial x}  
+ \left( 8 u^2_x  + 8  T - 4  \right) \beta c \tau_1 \frac{\partial E_p}{\partial y} \nonumber \\
+ 8 u_y \beta c \tau_2 E_p  \frac{\partial u_j}{\partial x_j}
- 8 \beta  \tilde{\mu} \left( u_y \tilde{\sigma}_{x,x} + u_x \left( \tilde{\sigma}_{x,y} + \tilde{\sigma}_{y,x} \right) \right)
- 4 \beta \tilde{\kappa} \frac{\partial T}{\partial y}
+H.O.
\end{align}
It is because 
\begin{align}
\int \delta f^{eq} \left( 8 v^2_x v_y - 4 v_y \right)  d v_x d v_y 
=- \beta \frac{\tilde{\mu} }{\rho T^2 } \frac{\partial u_l}{\partial x_k} \int 8 v^2_x v_y f^{eq} \left\{ \left(v_l -u_l \right) \left(v_k -u_k \right)  - \frac{\left( v-u \right)^2}{2} \delta_{l,k} \right\}  dv_x dv_y \nonumber \\
- \beta  \frac{ \tilde{\kappa} }{4 \rho T^3 }  \frac{\partial T}{\partial x_k}  \int  8 v^2_x v_y  f^{eq} \left\{ \left( v -u \right)^2 -4  T \right\} \left( v_k - u_k \right)dv_x dv_y \nonumber \\
= - 8 \beta  \tilde{\mu} \left( u_y \tilde{\sigma}_{x,x} + u_x \tilde{\sigma}_{x,y} + u_x \tilde{\sigma}_{y,x} \right)
- 4 \beta  \tilde{\kappa} \frac{\partial T}{\partial y}
\end{align}
using Eq.~(\ref{delta_feq}), Eq.~(\ref{feq_moment1}), and Eq.~(\ref{feq_moment2}).

For $d^{eq}_{4,0}$, using Eq.~(\ref{u_renormalization}) and Eq.~(\ref{T_renormalization}), we obtain,
\begin{align}
d^{eq}_{4,0}= \int \int \tilde{f}^{eq} \left( 16  v^4_x  - 48 v^2_x +12 \right) d v_x d v_y 
=48 \rho \left( T^2  + 2 u^2_x T  + u^4_x \right) - 48 \rho  T -48 \rho u_x^2 +12 \rho  + H.O.
\end{align}

For $d^{eq}_{3,1}$, using Eq.~(\ref{u_renormalization}) and Eq.~(\ref{T_renormalization}), we obtain,
\begin{align}
d^{eq}_{3,1}= \int \int \tilde{f}^{eq} \left( 16  v^3_x v_y  - 24 v_x v_y \right) d v_x d v_y 
=16 \left( \rho u^3_x + 3 \rho u_x T  \right) u_y - 24\rho u_x u_y + H.O.
\end{align}

For $d^{eq}_{2,2}$, using Eq.~(\ref{u_renormalization}) and Eq.~(\ref{T_renormalization}), we obtain,
\begin{align}
d^{eq}_{2,2}= \int \int \tilde{f}^{eq} \left( 16  v^2_x v^2_y  - 8 \left( v^2_x + v^2_y \right) +4 \right) d v_x d v_y \nonumber \\
= 16 \rho \left( T  + u^2_x \right) \left( T  + u^2_y \right) -16 \rho  T - 8 \rho \left( u^2_x + u^2_y \right)  + 4 \rho + H.O.
\end{align}

The results are summarized in Table.~\ref{tab:moment_new_eq}.

\begin{table*}[htbp]
    \caption{Symbolical forms of the projected equilibrium state in Eq.~(\ref{new_eq_def}) into the Hermite space for each rank of Hermite polynomials (raw) and each $\epsilon$ order (column).}
        \begin{center}
            \tabcolsep = 0.15cm
	\begin{tabular}{c | c | c | c } \hline
	 & zeroth   & first     &  second      \\ \hline
          $ d^{eq}_{0,0}$ & $\rho$    &    0 &  0            \\  & & &  \\
          $ d^{eq}_{1,0}$      &    $2 \rho u_x$ &  $2 \beta c \tau_1 \frac{\partial E_P}{\partial x}$       &    $0$      \\  & & &  \\
          $ d^{eq}_{2,0}$      &    $4 \left(  \rho T + \rho u_x^2 \right)-2 \rho$  & $4 \beta c \tau_2 E_P \frac{\partial u_j}{\partial x_j}+ 8 \beta c \tau_1 u_x \frac{\partial E_P}{\partial x}  $    &    $ - 2 \frac{ \beta c^2 \tau_2 \tau_1}{\rho} \left| \frac{\partial  E_p}{\partial x_i} \right|^2$           \\ 
         &  & $-2 \beta \tilde{\mu} \left( \tilde{\sigma}_{x,x} - \tilde{\sigma}_{y,y} \right)$    & $  
+  4 \frac{\beta^2 c^2 \tau^2_1}{\rho} \left( \frac{\partial E_p}{\partial x}  \right)^2$ \\ 
           & & $  $ &  \\& & &  \\                   
           	$ d^{eq}_{1,1}$      & $4 \rho u_x u_y$ & $4 \beta c \tau_1 \left( u_x \frac{\partial E_p}{\partial y} + u_y \frac{\partial E_p}{\partial x} \right)$   & $- $ \\ 
           	   & & $-4 \beta  \tilde{\mu} \tilde{\sigma}_{x,y}$ &  \\ & & &  \\
            	$ d^{eq}_{3,0}$      & $8 \rho u_x^3 + 24 u_x  \rho T - 12 \rho u_x$ & $\beta c \tau_1 \left( 24 u_x^2 + 24  T -12 \right) \frac{\partial E_p}{\partial x}$ & - \\
            	 & &  $+24 u_x \beta c \tau_2 E_p \frac{\partial u_j}{\partial x_j}$ & \\ 
            	 & & $-12 \beta  \tilde{\mu} u_x \left( \tilde{\sigma}_{x,x} - \tilde{\sigma}_{y,y} \right)-12 \beta  \tilde{\kappa} \frac{ \partial T}{\partial x}$ & \\
            	 & & &  \\
            	$ d^{eq}_{2,1}$      &  	$8 \rho u_x^2 u_y + 8 u_y  \rho T - 4 \rho u_y$  & $ 16 u_x u_y \beta c \tau_1 \frac{\partial E_p}{\partial x}$ & - \\         
            	&   	&  $+\beta c \tau_1 \left( 8 u_x^2 -4 +8  T \right) \frac{\partial E_p}{\partial y} $ & - \\    	
             &   	&  $+ 8 \beta c \tau_2 u_y E_p \frac{\partial u_j}{\partial x_j}$ & - \\ 
             &    &  $-8 \beta  \tilde{\mu} \left( u_y \tilde{\sigma}_{x,x} + u_x \left( \tilde{\sigma}_{x,y} + \tilde{\sigma}_{y,x} \right)  \right) -4 \beta  \tilde{\kappa} \frac{\partial T}{\partial y}$ & \\
             & & &  \\
             	$ d^{eq}_{4,0}$      & $48 \rho \left( T^2 + 2 u_x^2 T  + u_x^4  \right. $ & - & - \\ 
             	& $\left. -  T - u_x^2 +\frac{1}{4} \right) $ & &  \\ & & &  \\
                  	$ d^{eq}_{3,1}$ & $16 \rho u_y \left(  u_x^3 + 3 u_x T  - \frac{3}{2} u_x \right) $ & - & - \\ & & &  \\
                  	$ d^{eq}_{2,2}$ & $16 \rho \left\{ \left( T  + u_x^2 \right) \left( T  + u_y^2 \right)  \right.$ & - & - \\
                  	 & $\left. - T -\frac{1}{2} \left( u_x^2 + u_y^2 \right) + \frac{1}{4} \right\}$ & & \\ & & &  \\   	 
           \hline
    \end{tabular}
    \label{tab:moment_new_eq}
  \end{center}
\end{table*}

\section{Derivation of the continuity equation}
\label{deriv_cont}

Where $\left\{ l, k \right\}= \left\{ 0,0 \right\}$, Eq.~(\ref{first-order_D}) is
\begin{eqnarray}
\label{Eq_continuity}
\frac{\partial d^{(0)}_{0,0}}{\partial t_{1}} + c \frac{\partial \mathcal{D}^{(0)}_{j,0,0} }{\partial x_{1, j}}= - \frac{d^{(1)}_{0,0}-d^{(eq,1)}_{0,0}}{\tau_{0}}.
\end{eqnarray}
Because we defined $d_{0,0}=d^{eq}_{0,0}$ and $d^{(eq,p)}_{0,0}=0$ for $p>0$, $d^{(p)}_{0,0}=0$ for $p > 0$.
Using this fact and Table.~\ref{tab:moment_new_eq}, Eq.~(\ref{Eq_continuity}) becomes,
\begin{eqnarray}
\label{continuity_eq_firstod}
\frac{\partial \rho}{\partial t_1} + c  \frac{\partial \rho u_j}{\partial x_{1, j}} =0.
\end{eqnarray}
which is the continuity equation in the $\epsilon$ order.
It is because
\begin{eqnarray}
\mathcal{D}^{(0)}_{x,0,0}=\frac{d^{(0)}_{1,0}}{2}=\frac{d^{(eq,0)}_{1,0}}{2}, \nonumber \\
\mathcal{D}^{(0)}_{y,0,0}=\frac{d^{(0)}_{0,1}}{2} = \frac{d^{(eq,0)}_{0,1}}{2}.
\end{eqnarray}

Where $\left\{ l, k \right\}= \left\{ 0,0 \right\}$, Eq.~(\ref{sec-order-last}) is
\begin{eqnarray}
\frac{\partial d^{(0)}_{0,0}}{\partial t_{2}} + \frac{\partial d^{(1)}_{0,0}}{\partial t_{1}} 
+ c  \frac{\partial \mathcal{D}^{(eq,1)}_{j,0,0}}{\partial x_{1,j}}
-c \left( \frac{\partial^2 \tilde{\mathcal{D}}_{j,0,0} [d^{(0)}]  }{\partial t_1 \partial x_{1, j}}   + c  \frac{\partial^2 \tilde{\mathcal{D}}_{i,0,0} [\mathcal{D}^{(0)}_{j,0,0}]  }{\partial x_{1, i} \partial x_{1, j}}  \right)
= - \frac{d^{(2)}_{0,0}-d^{(eq,2)}_{0,0}}{\tau_{0}}.
\end{eqnarray}
Using $d^{(p)}_{0,0}=d^{(eq,p)}_{0,0}=0$ for $p > 0$ and Table.~\ref{tab:moment_new_eq}, it can be written as,
\begin{eqnarray}
\label{HO_continuity_deriv}
\frac{\partial \rho}{\partial t_{2}}  
+ c   \frac{\partial \mathcal{D}^{(eq,1)}_{j,0,0}}{\partial x_{1, j}}
-c \left(  \frac{\partial^2 \tilde{\mathcal{D}}_{j,0,0} [d^{(0)}]  }{\partial t_1 \partial x_{1,j}}   + c \frac{\partial^2 \tilde{\mathcal{D}}_{i,0,0} [\mathcal{D}^{(0)}_{j,0,0}]  }{\partial x_{1,i} \partial x_{1,j}} \right)
=0.
\end{eqnarray}
The left second term in Eq.~(\ref{HO_continuity_deriv}) can be written as,
\begin{eqnarray}
\label{ccterm1}
\frac{\partial \mathcal{D}^{(eq,1)}_{j,0,0}}{\partial x_{1,j}}=\frac{\partial}{\partial x_1} \left(  \frac{d^{(eq,1)}_{1,0}}{2} \right)  + \frac{\partial}{\partial y_1} \left(  \frac{d^{(eq,1)}_{0,1}}{2} \right) 
= \underbrace{ \beta c \tau_1 \frac{\partial^2 E_p} {\partial x_{1,j} \partial x_{1,j} }}_{cterm1}
\end{eqnarray}
The left third term in Eq.~(\ref{HO_continuity_deriv}) can be written as,
\begin{eqnarray}
\label{ccterm2}
 \frac{\partial^2 \tilde{\mathcal{D}}_{j,0,0} [d^{(0)}]  }{\partial t_1 \partial x_{1,j}} = 
\tau_1  \frac{\partial}{\partial t_1} \left(  \frac{\partial}{\partial x_1} \left(  \frac{d^{(eq,0)}_{1,0}}{2} \right) + \frac{\partial}{\partial y_1} \left(  \frac{d^{(eq,0)}_{0,1}}{2} \right)  \right), \nonumber \\
=\tau_1 c \left\{ \frac{\partial}{\partial x_{1,l}} \left(\underbrace{  -\frac{\partial \rho u_j u_l}{\partial x_{1,j}}}_{cterm2} -\underbrace{  \frac{\partial P}{\partial x_{1,l}}}_{cterm3}
\right)  \right\}.
\end{eqnarray}
Here we used the Navier-Stokes equation in the $\epsilon$ order, Eq.~(\ref{NS-leading-order}).
The left fourth term in Eq.~(\ref{HO_continuity_deriv}) can be written as,
\begin{eqnarray}
\label{ccterm3}
 \frac{\partial^2 \tilde{\mathcal{D}}_{i,0,0} [\mathcal{D}^{(0)}_{j,0,0}]  }{\partial x_{1,i} \partial x_{1,j}} =
 -c \frac{\tau_1}{2} \left\{\frac{\partial^2}{\partial x_1^2}\left( \frac{d^{(eq,0)}_{2,0}}{2} + d^{(eq,0)}_{0,0} \right) +\frac{\partial^2}{\partial x_1 \partial y_1} d^{(eq,0)}_{1,1}
 +\frac{\partial^2}{\partial y_1^2}\left( \frac{d^{(eq,0)}_{0,2}}{2} + d^{(eq,0)}_{0,0} \right) 
  \right\}, \nonumber \\
= -c \tau_1  \left\{
\underbrace{ \frac{\partial^2  \rho T}{\partial x_{1,j} \partial x_{1,j}}  }_{cterm4}
+\underbrace{ \frac{\partial^2 \rho u^2_x}{\partial x_1^2}  +  \frac{\partial^2 \rho u^2_y}{\partial y_1^2} }_{cterm5}
+\underbrace{ 2 \frac{\partial^2 \rho u_x u_y}{\partial x_1 \partial y_1} }_{cterm6}
\right\}.
\end{eqnarray}
Combining Eq.~(\ref{ccterm1}), Eq.~(\ref{ccterm2}), and Eq.~(\ref{ccterm3}),  we see 
\begin{eqnarray}
[cterm1]+[cterm3]+[cterm4]=0,
\end{eqnarray}
where $\beta=1$ or $\beta=0$ in the case of the ideal monoatomic gas. In addition,
\begin{eqnarray}
[cterm2]+[cterm5]+[cterm6]=0.
\end{eqnarray}
As a result, Eq.~(\ref{HO_continuity_deriv}) becomes,
\begin{eqnarray}
\label{continuity_eq_secondod}
\frac{\partial \rho}{\partial t_{2}}  =0.
\end{eqnarray}
which is the continuity equation in the $\epsilon^2$ order.

Assembling the temporal derivatives up to $\epsilon^2$ order, we derive the continuity equation,
\begin{eqnarray}
\frac{\partial \rho}{\partial t} + c  \frac{\partial \rho u_j}{\partial x_j} =0,
\end{eqnarray}
with the second-order accuracy of $\epsilon$.


\section{Derivation of the Navier-Stokes equation}
\label{deriv_NS}

Where $\left\{ l, k \right\}= \left\{ 1,0 \right\}$, Eq.~(\ref{first-order_D}) is
\begin{eqnarray}
\label{Eq_NS-deriv}
\frac{\partial d^{(0)}_{1,0}}{\partial t_{1}} + c \frac{\partial \mathcal{D}^{(0)}_{j,1,0} }{\partial x_{1,j}}= - \frac{d^{(1)}_{1,0}-d^{(eq,1)}_{1,0}}{\tau_{1}}.
\end{eqnarray}
Because we define $d_{1,0}=d^{(eq,0)}_{1,0}= 2 \rho u_x$, $d^{(p)}_{1,0}=0$ for $p > 0$.
In addition, using Table.~\ref{tab:moment_new_eq}, Eq.~(\ref{Eq_NS-deriv}) can be written as,
\begin{eqnarray}
\label{Eq_NS-deriv2}
2 \frac{\partial \rho u_x}{\partial t_{1}} + c \frac{\partial \mathcal{D}^{(0)}_{j,1,0} }{\partial x_{1,j}}=  2 \beta c \frac{\partial E_p}{\partial x_1}.
\end{eqnarray}
Moreover, because
\begin{eqnarray}
\frac{\partial \mathcal{D}^{(0)}_{j,1,0} }{\partial x_{1,j}}= \frac{\partial}{\partial x_1} \left( \frac{d^{(0)}_{2,0}}{2} + d^{(0)}_{0,0} \right) + \frac{\partial}{\partial y_1} \left( \frac{d^{(0)}_{1,1} }{2} \right), \nonumber \\
=\frac{\partial}{\partial x_1} \left(  2 \left(  \rho T +  \rho u^2_x \right) \right)
 +\frac{\partial}{\partial y_1} \left(2 \rho u_x u_y \right),
\end{eqnarray}
Eq.~(\ref{Eq_NS-deriv2}) can be written as,
\begin{eqnarray}
\frac{\partial \rho u_x}{\partial t_{1}} +c  \frac{\partial \rho u_x u_j}{\partial x_{1,j}} = -c \frac{\partial}{\partial x_1} \left(  \rho T - \beta E_p \right). 
\end{eqnarray}
Where $\beta=1$ or in the case of the ideal monoatomic gas at $\beta=0$, it becomes
\begin{eqnarray}
\label{NS-leading-order}
\frac{\partial \rho u_x}{\partial t_{1}} +c  \frac{\partial \rho u_x u_j}{\partial x_{1,j}} = -c \frac{\partial P}{\partial x_1}, 
\end{eqnarray}
which is the Navier-Stokes equation in the $\epsilon$ order.

Where $\left\{ l, k \right\}= \left\{ 1,0 \right\}$, Eq.~(\ref{sec-order-last}) is
\begin{eqnarray}
\frac{\partial d^{(0)}_{1,0}}{\partial t_{2}} + \frac{\partial d^{(1)}_{1,0}}{\partial t_{1}} 
+ c  \frac{\partial \mathcal{D}^{(eq,1)}_{j,1,0}}{\partial x_{1,j}}
-c \left( \frac{\partial^2 \tilde{\mathcal{D}}_{j,1,0} [d^{(0)}]  }{\partial t_1 \partial x_{1,j}}   + c  \frac{\partial^2 \tilde{\mathcal{D}}_{i,1,0} [\mathcal{D}^{(0)}_{j,1,0}]  }{\partial x_{1,i} \partial x_{1,j}}  \right)
= - \frac{d^{(2)}_{1,0}-d^{(eq,2)}_{1,0}}{\tau_{1}}.
\end{eqnarray}
Using $d^{(p)}_{1,0}=0$ for $p > 0$ and Table.~\ref{tab:moment_new_eq}, it can be written as,
\begin{eqnarray}
\label{HO_NS_deriv}
\frac{2}{c} \frac{\partial \rho u_x }{\partial t_{2}}  =
-  \frac{\partial \mathcal{D}^{(eq,1)}_{j,1,0}}{\partial x_{1,j}}
+  \frac{\partial^2 \tilde{\mathcal{D}}_{j,1,0} [d^{(0)}]  }{\partial t_1 \partial x_{1,j}}   + c  \frac{\partial^2 \tilde{\mathcal{D}}_{i,1,0} [\mathcal{D}^{(0)}_{j,1,0}]  }{\partial x_{1,i} \partial x_{1,j}}.
\end{eqnarray}

The first term in the right-hand side can be written as,
\begin{eqnarray}
\label{first_rhs_NSHO}
-\frac{\partial \mathcal{D}^{(eq,1)}_{j,1,0}}{\partial x_{1,j}}
=-\frac{\partial}{\partial x_1} \left(  \frac{d^{(eq,1)}_{2,0}}{2} - d^{(eq,1)}_{0,0} \right)  - \frac{\partial}{\partial y_1} \left(  \frac{d^{(eq,1)}_{1,1}}{2} \right), \nonumber \\
=- \frac{\partial}{\partial x_1} \left\{  2 \beta c \tau_2 E_P \frac{\partial u_j}{\partial x_{1,j}} + 4 \beta c \tau_1 u_x \frac{\partial E_P}{\partial x_1}  \right\}
- \frac{\partial}{\partial y_1} \left\{  2 \beta c \tau_1 \left( u_x \frac{\partial E_p}{\partial y_1} + u_y \frac{\partial E_p}{\partial x_1} \right) \right\} \nonumber \\
+ \beta \left( \frac{\partial \tilde{\mu} \left( \tilde{\sigma}_{x,x}- \tilde{\sigma}_{y,y} \right) }{\partial x_1} + 2 \frac{\partial \tilde{\mu} \tilde{\sigma}_{x,y}}{\partial y_1} \right) \nonumber \\
=-2 \beta \tau c \left[ \frac{\partial}{\partial x_1} \left\{ 2 \frac{\partial E_p u_x}{\partial x_1}  - E_p \left( \frac{\partial u_x}{\partial x_1} -  \frac{\partial u_y}{\partial y_1}  \right) \right\} \right. \nonumber \\
\left. + \frac{\partial}{\partial y_1} \left\{ \frac{\partial \rho T u_x }{\partial y_1}   -\rho T   \frac{\partial u_x}{\partial y_1} +  \frac{\partial \rho T u_y }{\partial x_1}  -\rho T   \frac{\partial u_y}{\partial x_1}  - u_x \frac{\partial P}{\partial y_1} - u_y   \frac{\partial P}{\partial x_1} \right\} \right] \nonumber \\
+ 2 \beta  \left( \frac{\partial \sigma_{1,x,j}}{\partial x_{1,j}} -c \frac{\partial \Pi_{1,x,j}}{\partial x_{1,j}} \right) \nonumber \\
= -2 \beta \tau c \left[ 2 \frac{\partial^2 E_p u_x}{\partial x_1^2}  
+ \frac{\partial }{\partial x_1} \left( \left(- \rho T + P \right) \left( \frac{\partial u_x}{\partial x_1} -  \frac{\partial u_y}{\partial y_1}  \right) \right)
+\frac{\partial^2 \rho T u_x }{\partial y_1^2}  \right. \nonumber \\
\left. +\frac{\partial^2 \rho T u_y }{\partial x_1 \partial y_1} 
-\frac{\partial}{\partial y_1} \left( \frac{\partial P}{\partial x_1} u_y + \frac{\partial P}{\partial y_1} u_x \right)
+ \frac{\partial }{\partial y_1} \left( \left(- \rho T \right) \left( \frac{\partial u_x}{\partial y_1} + \frac{\partial u_y}{\partial x_1}  \right) \right)
\right] \nonumber \\
+ 2 \beta  \left( \frac{\partial \sigma_{1,x,j}}{\partial x_{1,j}} -c \frac{\partial \Pi_{1,x,j}}{\partial x_{1,j}} \right) \nonumber \\
\label{first_HO_term_2}
=-2 \beta \tau c \left[ \underbrace{ 2 \frac{\partial^2 E_p u_x}{\partial x_1^2} }_{NSHO-A}
+\underbrace{   \frac{\partial }{\partial x_1} \left( P \left( \frac{\partial u_x}{\partial x_1} -  \frac{\partial u_y}{\partial y_1}  \right) \right)}_{NSHO-B}
+\underbrace{\frac{\partial^2 \rho T u_x }{\partial y_1^2} }_{NSHO-C} \right. \nonumber \\
\left. +\underbrace{\frac{\partial^2 \rho T u_y }{\partial x_1 \partial y_1} }_{NSHO-D}
-\underbrace{\frac{\partial}{\partial y_1} \left( \frac{\partial P}{\partial x_1} u_y + \frac{\partial P}{\partial y_1} u_x \right)}_{NSHO-E}
 \right] + 2 \beta  \frac{\partial \sigma_{1,x,j}}{\partial x_{1,j}}
\end{eqnarray}
In the third equation, we assume $\tau_1 = \tau_2 = \tau$.

The second term in the right-hand side of Eq.~(\ref{HO_NS_deriv}) can be written as,
\begin{eqnarray}
\label{third_term_1}
 \frac{\partial^2 \tilde{\mathcal{D}}_{j,0,0} [d^{(0)}]  }{\partial t_1 \partial x_{1,j}} = 
 \frac{\partial}{\partial t_1} \left(   \frac{\partial}{\partial x_1} \left(  \tau_2 \frac{d^{(0)}_{2,0}}{2} + \tau_0 d^{(0)}_{0,0} \right) + \frac{\partial}{\partial y_1} \left( \tau_2 \frac{d^{(0)}_{1,1}}{2} \right)  \right), \nonumber \\
=\frac{\tau_2}{2} \frac{\partial^2  \left( 4 \left(  \rho T + \rho u^2_x \right) - 2 \rho \right) }{\partial t_1 \partial x_1} + \frac{\tau_2}{2} \frac{\partial^2 \left( 4 \rho u_x u_y \right)}{\partial t_1 \partial y_1} + \tau_0 \frac{\partial^2 \rho}{\partial t_1 \partial x_1}.
\end{eqnarray}
If we use the continuity equation in the $\epsilon$ order, Eq.~(\ref{continuity_eq_firstod}), and the Navier-Stokes equation in the $\epsilon$ order, Eq.~(\ref{NS-leading-order}), we can derive,
\begin{eqnarray}
\label{rho_u_temporal_deriv}
\rho \frac{\partial u_x}{\partial t_1} = -u_x \frac{\partial \rho}{\partial t_1} - c \frac{\partial \rho u_x u_j}{\partial x_{1,j}}-c \frac{\partial P}{\partial x_1}, \nonumber \\
=c u_x \frac{\partial \rho u_j}{\partial x_{1,j}}-c \frac{\partial \rho u_x u_j}{\partial x_{1,j}}-c \frac{\partial P}{\partial x_1}, \nonumber \\
=-c \left( \rho u_j \frac{\partial u_x}{\partial x_{1,j}} \right) -c \frac{\partial P}{\partial x_1}.
\end{eqnarray}
Using it, Eq.~(\ref{continuity_eq_firstod}), and Eq.~(\ref{NS-leading-order}), we can derive,
\begin{eqnarray}
\label{rho_u2_temporal_deriv}
\frac{\partial \rho u^2_x}{\partial t_1}=u^2_x \frac{\partial \rho}{\partial t_1} + 2 \rho u_x \frac{\partial u_x}{\partial t_1}, \nonumber \\
=-u^2_x c \left(   \frac{\partial \rho u_j}{\partial x_{1,j}}  \right) -2 c \left( \rho u_x u_j \frac{\partial u_x}{\partial x_{1,j}} \right) -2c u_x  \frac{\partial P}{\partial x_1}, \nonumber \\
= -c \left(\frac{\partial \rho u^2_x u_j}{\partial x_{1,j}}+2 u_x \frac{\partial P}{\partial x_1} \right).
\end{eqnarray}
and
\begin{eqnarray}
\label{rho_uxuy_temporal_deriv}
\frac{\partial \rho u_x u_y}{\partial t_1}=\rho u_x \frac{\partial u_y}{\partial t_1} + u_y \frac{\partial \rho u_x}{\partial t_1}, \nonumber \\
=- c \left( \rho u_x u_y \frac{\partial u_y}{\partial y_1} + \rho u_x u_y \frac{\partial u_y}{\partial x_1} \right) - c u_x \frac{\partial P}{\partial y_1} - c \left( u_y \frac{\partial \rho u^2_x}{\partial x_1} + u_y \frac{\partial \rho u_x u_y}{\partial y_1} - u_y \frac{\partial P}{\partial x_1}\right), \nonumber \\
= -c \left(\frac{\partial \rho u_x u_y u_j}{\partial x_{1,j}} + u_x \frac{\partial P}{\partial y_1} +   u_y \frac{\partial P}{\partial x_1}   \right).
\end{eqnarray}
Applying Eq.~(\ref{rho_u2_temporal_deriv}) and Eq.~(\ref{rho_uxuy_temporal_deriv}) to Eq.~(\ref{third_term_1}), where $\tau_0 = \tau_2 =\tau$ we obtain,
\begin{eqnarray}
\label{third_term_2}
\frac{\partial^2 \tilde{\mathcal{D}}_{j,0,0} [d^{(0)}]  }{\partial t_1 \partial x_{1,j}} = \nonumber \\
-c \frac{\tau}{2} 4 \left[ \frac{\partial}{\partial x_1} \left\{  \frac{\partial  \rho u_j T}{\partial x_{1,j}}  +P \frac{\partial u_j}{\partial x_{1,j}}  \right\}
+\frac{\partial}{\partial x_1} \left\{  \left( \frac{\partial \rho u^2_x u_j}{\partial x_{1,j}} +2 u_x \frac{\partial P}{\partial x_1} \right) \right\} \right. \nonumber \\
\left.
+\frac{\partial}{\partial y_1} \left\{   \left( \frac{\partial \rho u_x u_y u_j}{\partial x_{1,j}} + u_x \frac{\partial P}{\partial y_1} +   u_y \frac{\partial P}{\partial x_1} \right) \right\}
\right], \nonumber \\
= -2 c \tau \left[ 
\underbrace{\frac{\partial \left(  \rho T +2P \right) u_x}{\partial x_1^2}}_{NSHO-A}
+\underbrace{\frac{\partial^2  \rho u_y T}{\partial x_1 \partial y_1}}_{NSHO-D}
+\underbrace{\frac{\partial}{\partial x_1} \left( P \left( -\frac{\partial u_x}{\partial x_1} + \frac{\partial u_y}{\partial y_1} \right)  \right)}_{NSHO-B}
+\underbrace{ \frac{\partial}{\partial y_1} \left(u_x \frac{\partial P}{\partial y_1} + u_y \frac{\partial P}{\partial x_1} \right) }_{NSHO-E} \right. \nonumber \\
\left.
+\underbrace{\frac{\partial^2 \rho u^3_x}{\partial x_1^2} +2 \frac{\partial^2 \rho u^2_x u_y}{\partial x_1 \partial y_1} + \frac{\partial^2 \rho u_x u^2_y}{\partial y_1^2}
}_{NSHO-F} 
\right]. \nonumber \\
%
\end{eqnarray}
Here we used Eq.~(\ref{continuity_eq_firstod}), Eq.~(\ref{NS-leading-order}), the heat-transfer equation in the $\epsilon$ order, Eq.~(\ref{Eq_thermal-LO}).

The third term in the right-hand side of Eq.~(\ref{HO_NS_deriv}) can be written as, 
\begin{eqnarray}
\label{fourth_term_1}
c  \frac{\partial^2 \tilde{\mathcal{D}}_{i,1,0} [\mathcal{D}^{(0)}_{j,1,0}]  }{\partial x_{1,i} \partial x_j}  \nonumber \\
= c \left[ \frac{\partial^2}{\partial x_1^2} \left( \frac{\tau_2}{2} \left( \frac{d^{(0)}_{3,0}}{2} + 2 d^{(0)}_{1,0} \right) + \frac{d^{(0)}_{1,0}}{2} \tau_0 \right) + \frac{\partial^2}{\partial y_1^2} \left(\frac{\tau_2}{2} \left( \frac{d^{(0)}_{1,2}}{2} + d^{(0)}_{1,0} \right) \right) \right. \nonumber \\
\left.
+\frac{\partial^2}{\partial x \partial y} \left( \frac{\tau_2}{2} \left( \frac{d^{(0)}_{2,1}}{2} \right) + \tau_0 \frac{d^{(0)}_{0,1}}{2} + \frac{\tau_2}{2} \left( \frac{d^{(0)}_{2,1}}{2} + d^{(0)}_{0,1} \right) \right)
\right], \nonumber \\
= c \tau \left[  \frac{\partial^2}{\partial x_1^2} \left(  2 \rho u^3+ 6 u_x  \rho T  \right)
+  \frac{\partial^2}{\partial y_1^2} \left( 2 \rho u_x u^2_y + 2 u_x  \rho T  \right) \right. \nonumber \\
\left.
+ \frac{\partial^2}{\partial x_1 \partial y_1} \left( 2 \rho u^2_x u_y + 4 u_y  \rho T + 2 \rho u^2_x u_y  \right)
\right], \nonumber \\
= c \tau  \left[ 
\underbrace{\frac{\partial^2}{\partial x_1^2} \left(6 \rho u_x  T \right)}_{NSHO-A}
+\underbrace{\frac{\partial^2}{\partial y_1^2} \left(2 \rho u_x  T \right)}_{NSHO-C}
+\underbrace{\frac{\partial^2}{\partial x_1 \partial y_1} \left(4 \rho u_y  T \right)}_{NSHO-D} \right. \nonumber \\
\left.
+\underbrace{\frac{\partial^2}{\partial x_1^2} \left(2 \rho u^3_x \right)
+\frac{\partial^2}{\partial y_1^2} \left(2 \rho u_x u^2_y \right)
+\frac{\partial^2}{\partial x_1 \partial y_1} \left(2 \rho u^2_x u_y + 2 \rho u_x u^2_y \right)
}_{NSHO-F}
\right],
\end{eqnarray}
where $\tau_0 = \tau_2 =\tau$.

Assembling Eq.~(\ref{first_HO_term_2}), Eq.~(\ref{third_term_2}), and  Eq.~(\ref{fourth_term_1}), we obtain,
\begin{eqnarray}
\left[ NSHO-A \right]= c \tau \frac{\partial^2}{\partial x_1^2} \left(- 4 \beta E_p u_x -2  \rho T u_x -  4P u_x  + 6 \rho u_x  T  \right) = 4 c \tau \left(1-\beta  \right) \frac{\partial^2}{\partial x_1^2} E_p u_x, \\
\left[ NSHO-B \right]= 2 c \tau \left( 1 - \beta \right) \frac{\partial}{\partial x_1} \left( P \left( \frac{\partial u_x}{\partial x_1} - \frac{\partial u_y}{\partial y_1} \right)  \right), \\
\left[ NSHO-C \right]= 2 c \tau \left( 1 - \beta \right) \frac{\partial^2}{\partial y_1^2} \left( \rho u_x  T \right), \\
\left[ NSHO-D \right]= 2 c \tau \left( 1 - \beta \right) \frac{\partial^2}{\partial x_1 \partial y_1} \left( \rho u_y  T \right), \\
\left[ NSHO-E \right]=2 c \tau \left( \beta-1 \right) \frac{\partial}{\partial y_1} \left(u_x \frac{\partial P}{\partial y_1} + u_y \frac{\partial P}{\partial x_1} \right), \\
\left[ NSHO-F \right]=\mathcal{O} \left( \frac{\partial^2 \rho u^3 }{\partial x^2} \right).
\end{eqnarray}

As a result, where $\beta=1$, all of them are canceled or negligible. Only the remaining term in the right-hand side is the last term of Eq.~(\ref{first_rhs_NSHO}). Therefore, Eq.~(\ref{HO_NS_deriv}) results in,
\begin{eqnarray}
\label{NS_secondod}
 \frac{\partial \rho u_x }{\partial t_{2}}  =c  \frac{\partial \sigma_{1,x,j}}{\partial x_{1,j}}.
\end{eqnarray}
which is the Navier-Stokes equation in the $\epsilon^2$ order.

On the other hand,  in the case of the ideal monoatomic gas, $P=\rho T$, at $\beta=0$, they becomes
\begin{eqnarray}
\left[ NSHO-A \right]=0, \\
\left[ NSHO-B \right]+\left[ NSHO-C \right] + \left[ NSHO-D \right] + \left[ NSHO-E \right]= \nonumber \\
2 c \tau \left\{ \frac{\partial}{\partial x_1} \left( P \left( \frac{\partial u_x}{\partial x_1} - \frac{\partial u_y}{\partial y_1} \right)  \right)
+\frac{\partial^2}{\partial y_1^2} \left( P u_x  \right)
+ \frac{\partial^2}{\partial x_1 \partial y_1} \left( P u_y \right)
-\frac{\partial}{\partial y_1} \left(u_x \frac{\partial P}{\partial y_1} + u_y \frac{\partial P}{\partial x_1} \right) 
\right\}, \nonumber \\
=2 c \tau \left\{
\frac{\partial}{\partial x_1} \left( P \left( \frac{\partial u_x}{\partial x_1} - \frac{\partial u_y}{\partial y_1} \right) \right) 
+\frac{\partial}{\partial y_1} \left( \frac{\partial P u_y}{\partial x_1} + \frac{\partial P u_x}{\partial y_1}  \right)
-\frac{\partial}{\partial y_1} \left(u_x \frac{\partial P}{\partial y_1} + u_y \frac{\partial P}{\partial x_1} \right) 
\right\}, \nonumber \\
=2 c \tau \left\{
\frac{\partial}{\partial x_1} \left( P \left( \frac{\partial u_x}{\partial x_1} - \frac{\partial u_y}{\partial y_1} \right) \right) 
+\frac{\partial}{\partial y_1} \left( P \left( \frac{\partial u_y}{\partial x_1} + \frac{\partial u_x}{\partial y_1} \right) \right) 
\right\}, \nonumber \\
= 2 c  \frac{\partial \Pi_{1,x,j}}{\partial x_{1,j}}.
\end{eqnarray}
As a result, Eq.~(\ref{HO_NS_deriv}) results in,
\begin{eqnarray}
 \frac{\partial \rho u_x }{\partial t_{2}}  = c^2   \frac{\partial \Pi_{1,x,j}}{\partial x_{1,j}},
\end{eqnarray}
which is equivalent to Eq.~(\ref{NS_secondod}) if the stress tensor $\sigma$ has shear viscosity of $c P \tau$ and zero bulk viscosity.

Assembling the temporal derivatives up to $\epsilon^2$ orders,  at $\beta=1$ we derive the Navier-Stokes equation,
\begin{eqnarray}
\frac{\partial \rho u_x}{\partial t} +c  \frac{\partial \rho u_x u_j}{\partial x_{j}} = -c \frac{\partial P}{\partial x} + c  \frac{\partial  \sigma_{x,j}}{\partial x_{j}}.
\end{eqnarray}
with the second-order accuracy of $\epsilon$.
In the case of the ideal monoatomic gas at $\beta=0$, the stress tensor $\sigma$ has shear viscosity of $c P \tau$ and zero bulk viscosity.

\section{Derivation of the heat transfer equation}
\label{deriv_heat}

The sum of Eq.~(\ref{first-order_D}) at $\left\{ l, k \right\}= \left\{ 2,0 \right\}$ and Eq.~(\ref{first-order_D}) at $\left\{ l, k \right\}= \left\{ 0,2 \right\}$ is,
\begin{eqnarray}
\label{Eq_thermal-deriv}
\frac{\partial d^{(0)}_{2,0} + d^{(0)}_{0,2} }{\partial t_{1}} + c \frac{\partial \mathcal{D}^{(0)}_{j,2,0} +  \mathcal{D}^{(0)}_{j,0,2}}{\partial x_{1,j}}= - \frac{d^{(1)}_{2,0}-d^{(eq,1)}_{2,0}+d^{(1)}_{0,2}-d^{(eq,1)}_{0,2}}{\tau_{2}}.
\end{eqnarray}
Because we define $d_{2,0}+d_{0,2} =d^{(eq,0)}_{2,0} + d^{(eq,0)}_{0,2}= 8 \rho T + 4 \rho u^2 - 4 \rho$, $d^{(p)}_{2,0} + d^{(p)}_{0,2}=0$ for $p > 0$.
In addition, using Table.~\ref{tab:moment_new_eq}, Eq.~(\ref{Eq_thermal-deriv}) becomes,
\begin{eqnarray}
\label{Eq_thermal-deriv2}
\frac{\partial \left( 8  \rho T + 4 \rho u^2 - 4 \rho \right)}{\partial t_{1}} 
+ c \frac{\partial }{\partial x_{1,j}} \left( 4 \rho u^3_j + 16 u_j  \rho T -4 \rho u_j + 4 \rho u_x u_y u_j \right) \nonumber \\
=  8 \beta c E_p \frac{\partial u_j}{\partial x_{1,j}}+8\beta c \frac{\tau_1}{\tau_2} u_j \frac{\partial E_p}{\partial x_{1,j}}.
\end{eqnarray}
If we apply  the continuity equation in the $\epsilon$ order, Eq.~(\ref{continuity_eq_firstod}), Eq.~(\ref{rho_u2_temporal_deriv}),  and its extension for $\frac{\partial \rho u^2_y}{\partial t_1}$, where $\tau_1=\tau_2$, it becomes,
\begin{eqnarray}
\label{Eq_thermal-deriv3}
8 \frac{\partial  \rho T}{\partial t_{1}}  - 8 c \left( \frac{\partial P u_j}{\partial x_{1,j}} - P \frac{\partial u_j}{\partial x_{1,j}} \right)
+ 16 c \frac{\partial u_j  \rho T}{\partial x_{1,j}} = 8 \beta c \frac{\partial E_p u_j}{\partial x_{1,j}}.
\end{eqnarray}
When $\beta=1$, Eq.~(\ref{Eq_thermal-deriv3}) can be arranged as,
\begin{eqnarray}
\label{Eq_thermal-LO}
\frac{\partial  \rho T}{\partial t_{1}}  + c \frac{\partial u_j  \rho T}{\partial x_{1,j}} = - c  P \frac{\partial u_j}{\partial x_{1,j}},
\end{eqnarray}
which is  the heat transfer equation in the $\epsilon$ order.

On the other hand, in the case of the ideal monoatomic gas, namely $P = \rho T$ at $\beta=0$, Eq.~(\ref{Eq_thermal-deriv3}) becomes,
\begin{eqnarray}
\label{Eq_thermal-LO2}
\frac{\partial  \rho T}{\partial t_{1}}  + c \frac{\partial u_j  \rho T}{\partial x_{1,j}} = - c  P \frac{\partial u_j}{\partial x_{1,j}},
\end{eqnarray}
which is equivalent to Eq.~(\ref{Eq_thermal-LO}) and the heat transfer equation in the $\epsilon$ order.

The sum of Eq.~(\ref{sec-order-last}) at $\left\{ l, k \right\}= \left\{ 2,0 \right\}$ and Eq.~(\ref{sec-order-last}) at $\left\{ l, k \right\}= \left\{ 0,2 \right\}$ is,
\begin{eqnarray}
\label{Eq_thermal-deriv-HO1}
\frac{\partial d^{(0)}_{2,0}+ d^{(0)}_{0,2}}{\partial t_{2}} 
+ c  \frac{\partial \mathcal{D}^{(eq,1)}_{j,2,0}+ \mathcal{D}^{(eq,1)}_{j,0,2}}{\partial x_{1,j}} \nonumber \\
-c \left( \frac{\partial^2 \tilde{\mathcal{D}}_{j,2,0} [d^{(0)}]  +\tilde{\mathcal{D}}_{j,0,2} [d^{(0)}]  }{\partial t_1 \partial x_{1,j}}   
+ c  \frac{\partial^2 \tilde{\mathcal{D}}_{i,2,0} [\mathcal{D}^{(0)}_{j,2,0}] + \tilde{\mathcal{D}}_{i,0,2} [\mathcal{D}^{(0)}_{j,0,2}] }{\partial x_{1,i} \partial x_{1,j}}  \right)
- \frac{d^{(eq,2)}_{2,0}+ d^{(eq,2)}_{0,2}}{\tau_{2}}=0.
\end{eqnarray}
Here we used $d^{(p)}_{2,0}+d^{(p)}_{0,2} =0$ for $p > 0$.

For convenience,  the following equation is derived using the continuity and Navier-Stokes equation in the $\epsilon^2$ order, Eq.~(\ref{continuity_eq_secondod}) and Eq.~(\ref{NS_secondod}),  
\begin{eqnarray}
\label{rho-u2_temporal2_deriv}
\frac{\partial \rho u^2}{\partial t_2}= 2 u_i \frac{\partial \rho u_i}{\partial t_2} = 2 c u_i \frac{\partial \sigma_{1,i,j}}{\partial x_{1,j}}.
\end{eqnarray}
In addition, using Eq.~(\ref{rho_u_temporal_deriv}), Eq.~(\ref{rho_u2_temporal_deriv}), the Navier-Stokes and the heat transfer equation in the $\epsilon$ order, Eq.~(\ref{NS-leading-order}), and Eq.~(\ref{Eq_thermal-LO}), we can derive following three equations,
\begin{eqnarray}
\label{rho_u3_temporal_deriv}
\frac{\partial \rho u^3_x}{\partial t_1} = u_x \frac{\partial \rho u^2_x}{\partial t_1} + \rho u^2_x \frac{\partial u_x}{\partial t_1} = -3 c u^2_x \frac{\partial P}{\partial x_1} + \mathcal{O} \left( \frac{\partial \rho u^4}{\partial x_1} \right),  \\
\label{rho_ux_uy2_temporal_deriv}
\frac{\partial \rho u_x u^2_y}{\partial t_1}= u_x \frac{\partial \rho u^2_y}{\partial t_1} + \rho u^2_y \frac{\partial u_x}{\partial t_1} = -2 c u_x u_y \frac{\partial P}{\partial y_1} - c u^2_y \frac{\partial P}{\partial x_1}+ \mathcal{O} \left( \frac{\partial \rho u^4}{\partial x_1} \right), \\
\label{cv_rho_t_ux_temporal_deriv}
\frac{\partial  \rho T u_x}{\partial t_1} = u_x \frac{\partial  \rho T}{\partial t_1} +  \rho T \frac{\partial u_x}{\partial t_1}= -c u_x \left(\frac{\partial  \rho T u_j}{\partial x_{1,j}} + P \frac{\partial u_j}{\partial x_{1,j}} \right) -   T c \left( \rho u_j \frac{\partial u_x}{\partial x_{1,j}} + \frac{\partial P}{\partial x_1} \right) .
\end{eqnarray}

Using Table.~\ref{tab:moment_new_eq}, the continuity equation in the $\epsilon^2$ order, Eq.~(\ref{continuity_eq_secondod}), and Eq.~(\ref{rho-u2_temporal2_deriv}), the first term in the left-hand side of  Eq.~(\ref{Eq_thermal-deriv-HO1}) becomes,
\begin{eqnarray}
\label{Eq_thermal-deriv-HO-left-first-1}
\frac{\partial d^{(0)}_{2,0}+ d^{(0)}_{0,2}}{\partial t_{2}} = 8 \frac{\partial  \rho T + 4 \rho u^2 - 4 \rho}{\partial t_2}= 8  \frac{\partial  \rho T}{\partial t_2} + 8 c u_i \frac{\partial \sigma_{i,j}}{\partial x_{1,j}}.
\end{eqnarray}

Using Table.~\ref{tab:moment_new_eq}, where $\tau_1=\tau_2 = \tau$, the second term in the left-hand side of  Eq.~(\ref{Eq_thermal-deriv-HO1}) becomes,
\begin{eqnarray}
\label{Eq_thermal-deriv-HO-left-first-2}
 c  \frac{\partial \mathcal{D}^{(eq,1)}_{j,2,0}+ \mathcal{D}^{(eq,1)}_{j,0,2}}{\partial x_{1,j}}
 = c \frac{\partial}{\partial x_1} \left( \frac{d^{(eq,1)}_{3,0} + 2 d^{(eq,1)}_{1,0} + d^{(eq,1)}_{1,2}}{2} \right)
   +\left( x \leftrightarrow y \right), \nonumber \\
   =\beta c^2 \tau \left\{ \frac{\partial}{\partial x_1}  \left[ \left(   4 u^2 + 8 u^2_x +16  T -4     \right) \frac{\partial E_p}{\partial x_1}
   +16 u_x E_p \frac{\partial u_j}{\partial x_{1,j}}
   \right]
    +\frac{\partial}{\partial x_1} \left( 8 u_x u_y \frac{\partial E_p}{\partial y_1}\right)
   \right\} \nonumber \\
   + \beta c \frac{\partial}{\partial x_1} \left( -6 \tilde{\mu} u_x \left( \tilde{\sigma}_{x,x} -  \tilde{\sigma}_{y,y} \right) - 6 \tilde{\kappa} \frac{\partial T}{\partial x} - 4 \tilde{\mu} \left( u_x \tilde{\sigma}_{y,y} + u_y \left(\tilde{\sigma}_{x,y} + \tilde{\sigma}_{y,x}  \right)  - 2 \tilde{\kappa} \frac{\partial T}{\partial x_1} \right)  \right)   
   +\left( x \leftrightarrow y \right). \nonumber \\
   =\beta c^2 \tau \left\{ \frac{\partial}{\partial x_1}  \left[ \left(   4 u^2 + 8 u^2_x +16  T -4     \right) \frac{\partial E_p}{\partial x_1}
   +16 u_x E_p \frac{\partial u_j}{\partial x_{1,j}}
   \right]
    +\frac{\partial}{\partial x_1} \left( 8 u_x u_y \frac{\partial E_p}{\partial y_1}\right)
   \right\} \nonumber \\
   - 8 \beta c  \frac{\partial}{\partial x_1} \left(  \tilde{\kappa} \frac{\partial T}{\partial x_1}\right)  - 8 \beta c \frac{\partial}{\partial x_1} \left(\tilde{\mu} u_l \tilde{\sigma}_{x,l} \right)+ \left( x \leftrightarrow y \right)
\end{eqnarray}
Here, $\left( x \leftrightarrow y \right)$ denotes the commuted terms in which $x$ and $y$ are exchanged from the entire former terms.

Using Table.~\ref{tab:moment_new_eq}, Eq.~(\ref{rho_u3_temporal_deriv}), Eq.~(\ref{rho_ux_uy2_temporal_deriv}), Eq.~(\ref{cv_rho_t_ux_temporal_deriv}), and the Navier-Stokes equation in the $\epsilon$ order, Eq.~(\ref{NS-leading-order}), the  third term  in the left-hand side of  Eq.~(\ref{Eq_thermal-deriv-HO1})  becomes,
\begin{eqnarray}
\label{Eq_thermal-deriv-HO-left-first-3}
-c \frac{\partial^2 \tilde{\mathcal{D}}_{j,2,0} [d^{(0)}]  +\tilde{\mathcal{D}}_{j,0,2} [d^{(0)}]  }{\partial t_1 \partial x_{1,j}}  \nonumber \\
= -c  \frac{\partial}{\partial t_1} \left[ \frac{\partial}{\partial x_1} \left(\tau_3 \frac{d^{(0)}_{3,0}}{2} + 2 \tau_1 d^{(0)}_{1,0} + \tau_3 \frac{d^{(0)}_{1,2}}{2} \right)
\right] + \left( x \leftrightarrow y \right), \nonumber \\
= - c  \frac{\partial}{\partial t_1} \left[ \frac{\partial}{\partial x_1} \left( \tau_3 \left(  4 \rho u^3_x + 16  \rho T u_x -8\rho u_x + 4 \rho u_x u^2_y \right)  + 4 \tau_1 \rho u_x  \right) \right] + \left( x \leftrightarrow y \right), \nonumber  \\
=-4 c^2 \frac{\partial}{\partial x_1} \left[ -3 \tau_3 u^2_x \frac{\partial P}{\partial x_1} 
-2 u_x u_y \tau_3 \frac{\partial P}{\partial y_1}
-u^2_y \tau_3 \frac{\partial P}{\partial x_1} \right. \nonumber \\
\left.
-\left( \underbrace{\frac{\partial P}{\partial x_1}}_{THHO-A}+
\underbrace{\frac{\partial \rho u^2_x}{\partial x_1}}_{THHO-B} +
\underbrace{\frac{\partial \rho u_x u_y}{\partial y_1}}_{THHO-C}
 \right) \frac{-8 \tau_3 + 4 \tau_1}{4} \right. \nonumber \\
 \left.
 - 4 \tau_3 u_x \left( \frac{\partial  \rho T u_j}{\partial x_{1,j}} + P \frac{\partial u_j}{\partial x_{1,j}} \right)
 -4  T \tau_3 \left( \rho u_j \frac{\partial u_x}{\partial x_{1,j}} + \frac{\partial P}{\partial x_1} \right)
\right]  \nonumber \\
+\mathcal{O}\left( \frac{\partial^2 \rho u^4}{\partial x^2_1} \right)+\left( x \leftrightarrow y \right).
\end{eqnarray}

Using Table.~\ref{tab:moment_new_eq}, the fourth term  in the left-hand side of  Eq.~(\ref{Eq_thermal-deriv-HO1})   becomes,
\begin{eqnarray}
\label{Eq_thermal-deriv-HO-left-first-4}
-c^2  \frac{\partial^2 \tilde{\mathcal{D}}_{i,2,0} [\mathcal{D}^{(0)}_{j,2,0}] + \tilde{\mathcal{D}}_{i,0,2} [\mathcal{D}^{(0)}_{j,0,2}] }{\partial x_{1,i} \partial x_{1,j}} \nonumber \\
= -c^2  \left\{
\frac{\partial^2}{\partial x_1^2} \left[ \tau_3 \frac{d^{(0)}_{4,0}}{4} +\left( \frac{3}{2}\tau_3 + \tau_1 \right) d^{(0)}_{2,0}
+ 2 \tau_1 d^{(0)}_{0,0} + \tau_3 \left( \frac{d^{(0)}_{2,2}}{4} + \frac{d^{(0)}_{0,2}}{2} \right)
  \right] \right. \nonumber \\ 
  \left.
 + \frac{\partial^2}{\partial x_1 \partial y_1} \left[
2 \left( \tau_1 + \tau_3 \right)  d^{(0)}_{1,1} + 2 \tau_3 \frac{d^{(0)}_{3,1}+d^{(0)}_{1,3}}{4}
    \right] 
\right\}+\left( x \leftrightarrow y \right), \nonumber  \\
=-c^2  \left\{
\frac{\partial^2}{\partial x_1^2} \left[ 
\underbrace{\rho  T \left(-8 \tau_3 +4 \tau_1 \right) }_{THHO-A}
+ \underbrace{\rho u^2_x \left( 4 \tau_1 -8 \tau_3 \right) }_{THHO-B}
+ \tau_3 \left( 28 \rho u^2_x T  + 4 \rho T  u^2_y + 16 \rho T^2  \right)
\right] \right. \nonumber \\
\left.
+ \frac{\partial^2}{\partial x_1 \partial y_1} \left[
\underbrace{\rho u_x u_y \left(-8 \tau_3 +4 \tau_1 \right) }_{THHO-C}
+ \tau_3 \left( 24 \rho u_x T  u_y \right)
\right]  \right\}
+ \mathcal{O} \left( \frac{\partial \rho u^4}{\partial x_1^2} \right)+ \left( x \leftrightarrow y \right).
\end{eqnarray}

Using Table.~\ref{tab:moment_new_eq}, the last term  in the left-hand side of  Eq.~(\ref{Eq_thermal-deriv-HO1})   becomes,
\begin{eqnarray}
\label{Eq_thermal-deriv-HO-left-last}
-\frac{d^{(eq,2)}_{2,0}+ d^{(eq,2)}_{0,2}}{\tau_{2}}= 
  0.
\end{eqnarray}

Next, we consider the sum of Eq.~(\ref{Eq_thermal-deriv-HO-left-first-3}) and Eq.~(\ref{Eq_thermal-deriv-HO-left-first-4}).
Because the sum for $[THHO-B]$ and $[THHO-C]$ are zero and the sum for $[THHO-A]$ is $c^2 \frac{\partial^2 E_p}{\partial x^2_1} \left( 8 \tau_3 - 4 \tau_1\right)$, we have,
\begin{eqnarray}
-c \frac{\partial^2 \tilde{\mathcal{D}}_{j,2,0} [d^{(0)}]  +\tilde{\mathcal{D}}_{j,0,2} [d^{(0)}]  }{\partial t_1 \partial x_{1,j}}-c^2  \frac{\partial^2 \tilde{\mathcal{D}}_{i,2,0} [\mathcal{D}^{(0)}_{j,2,0}] + \tilde{\mathcal{D}}_{i,0,2} [\mathcal{D}^{(0)}_{j,0,2}] }{\partial x_{1,i} \partial x_{1,j}} \nonumber \\
=-4 c^2 \frac{\partial}{\partial x_1} \left[ -3 \tau_3 u^2_x \frac{\partial P}{\partial x_1} 
-2 u_x u_y \tau_3 \frac{\partial P}{\partial y_1}
-u^2_y \tau_3 \frac{\partial P}{\partial x_1} 
+\frac{\partial E_p}{\partial x_1} \left( \frac{-8 \tau_3 + 4 \tau_1}{4} \right) \right. \nonumber \\
 \left.
 - 4 \tau_3 u_x \left( \frac{\partial  \rho T u_j}{\partial x_{1,j}} + P \frac{\partial u_j}{\partial x_{1,j}} \right)
 -4  T \tau_3 \left( \rho u_j \frac{\partial u_x}{\partial x_{1,j}} + \frac{\partial P}{\partial x_1} \right)
\right]  \nonumber \\
-c^2  \left\{
\frac{\partial^2}{\partial x_1^2} \left[ 
 \tau_3 \left( 28 \rho u^2_x T  + 4 \rho T  u^2_y + 16 \rho T^2  \right)
\right] 
+ \frac{\partial^2}{\partial x_1 \partial y_1} \left[
 \tau_3 \left( 24 \rho u_x T  u_y \right)
\right] 
\right\} \nonumber \\
 + \mathcal{O} \left( \frac{\partial \rho u^4}{\partial x_1^2} \right) +\left(x \leftrightarrow y \right),  \nonumber  \\
\label{sum_3-4}
=-c^2  \tau \frac{\partial}{\partial x_1} \left[ 
  \left(4 u^2 + 8 u^2_x + 16  T -4 \right) \frac{\partial E_p}{\partial x_1} 
 + 16 u_x E_p \frac{\partial u_j}{\partial x_{1,j}} \right.  \nonumber \\
 \left.
 +8u_x u_y \frac{\partial E_p}{\partial y_1} 
 + 8 \rho  T \left( u_x \frac{\partial u_x}{\partial x_1}- u_x \frac{\partial u_y}{\partial y_1}+   u_y \frac{\partial u_y}{\partial x_1}+u_y \frac{\partial u_x}{\partial y_1} \right) 
-16  T  \frac{\partial  \rho T}{\partial x_1}
\right]  - 16  c^2 \tau \frac{\partial^2 \rho T^2 }{\partial x_1^2}\nonumber \\
+ \mathcal{O} \left( \frac{\partial \rho u^4}{\partial x_1^2} \right) +\left(x \leftrightarrow y \right), \nonumber \\
\end{eqnarray}
where $\tau_3=\tau_1=\tau$.

Where $\beta=1$, the sum of Eq.~(\ref{Eq_thermal-deriv-HO-left-first-2}), Eq.~(\ref{Eq_thermal-deriv-HO-left-last}), and Eq.~(\ref{sum_3-4}) can be written as,
\begin{eqnarray}
\label{Heat_transfer_dum1}
c  \frac{\partial \mathcal{D}^{(eq,1)}_{j,2,0}+ \mathcal{D}^{(eq,1)}_{j,0,2}}{\partial x_{1,j}}-c \frac{\partial^2 \tilde{\mathcal{D}}_{j,2,0} [d^{(0)}]  +\tilde{\mathcal{D}}_{j,0,2} [d^{(0)}]  }{\partial t_1 \partial x_{1,j}}-c^2  \frac{\partial^2 \tilde{\mathcal{D}}_{i,2,0} [\mathcal{D}^{(0)}_{j,2,0}] + \tilde{\mathcal{D}}_{i,0,2} [\mathcal{D}^{(0)}_{j,0,2}] }{\partial x_{1,i} \partial x_{1,j}} -\frac{d^{(eq,2)}_{2,0}+ d^{(eq,2)}_{0,2}}{\tau_{2}} \nonumber \\
=-c^2  \tau \frac{\partial}{\partial x_1} \left[ 
 + 8 \rho  T \left( u_x \frac{\partial u_x}{\partial x_1}- u_x \frac{\partial u_y}{\partial y_1}+   u_y \frac{\partial u_y}{\partial x_1}+u_y \frac{\partial u_x}{\partial y_1} \right) 
 +16  \rho T  \frac{\partial  T}{\partial x_1}
\right] \nonumber \\
-8  c \frac{\partial }{\partial x_{1}} \left( \tilde{\kappa} \frac{\partial T}{\partial x_{1}} \right)
-8  c \frac{\partial \tilde{\mu} u_l \sigma_{l,x}}{\partial x_{1}}
 + \mathcal{O} \left( \frac{\partial \rho u^4}{\partial x_1^2} \right) +\left(x \leftrightarrow y \right), \nonumber \\
=-8c^2  \frac{\partial}{\partial x_{1,j}} \left[ \Pi_{1,i,j} u_i +2  \rho T \tau \frac{\partial  T}{\partial x_{1,j}} \right] 
-8  c \frac{\partial }{\partial x_{1,j}} \left( \tilde{\kappa} \frac{\partial T}{\partial x_{1,j}} \right)
-8  c \frac{\partial \tilde{\mu} u_l \sigma_{l,j}}{\partial x_{1,j}}
+ \mathcal{O} \left( \frac{\partial \rho u^4}{\partial x_1^2} \right). \nonumber \\
\end{eqnarray}
In the case of the monoatomic ideal gas at $\beta=0$, the sum of Eq.~(\ref{Eq_thermal-deriv-HO-left-first-2}) and Eq.~(\ref{sum_3-4}) can be written as,
\begin{eqnarray}
c  \frac{\partial \mathcal{D}^{(eq,1)}_{j,2,0}+ \mathcal{D}^{(eq,1)}_{j,0,2}}{\partial x_{1,j}}-c \frac{\partial^2 \tilde{\mathcal{D}}_{j,2,0} [d^{(0)}]  +\tilde{\mathcal{D}}_{j,0,2} [d^{(0)}]  }{\partial t_1 \partial x_{1,j}}-c^2  \frac{\partial^2 \tilde{\mathcal{D}}_{i,2,0} [\mathcal{D}^{(0)}_{j,2,0}] + \tilde{\mathcal{D}}_{i,0,2} [\mathcal{D}^{(0)}_{j,0,2}] }{\partial x_{1,i} \partial x_{1,j}}  -\frac{d^{(eq,2)}_{2,0}+ d^{(eq,2)}_{0,2}}{\tau_{2}}  \nonumber \\
=-c^2  \tau \frac{\partial}{\partial x_1} \left[ 
  8 \rho  T \left( u_x \frac{\partial u_x}{\partial x_1}- u_x \frac{\partial u_y}{\partial y_1}+   u_y \frac{\partial u_y}{\partial x_1}+u_y \frac{\partial u_x}{\partial y_1} \right) 
-16  T  \frac{\partial  \rho T}{\partial x_1}
\right]  - 16  c^2 \tau \frac{\partial^2 \rho T^2 }{\partial x_1^2}\nonumber \\
+ \mathcal{O} \left( \frac{\partial \rho u^4}{\partial x_1^2} \right) +\left(x \leftrightarrow y \right),  \nonumber \\
=-8c^2  \frac{\partial}{\partial x_{1,j}} \left[ \Pi_{1,i,j} u_i +2 P   \tau \frac{\partial T}{\partial x_{1,j}} \right]+ \mathcal{O} \left( \frac{\partial \rho u^4}{\partial x_1^2} \right), \nonumber  \\
\end{eqnarray}
which is equivalent to Eq.~(\ref{Heat_transfer_dum1}) because $\tilde{\kappa}=\tilde{\mu}=0$ and $P=\rho T$ in this condition.

Lastly, assembling the above equation and Eq.~(\ref{Eq_thermal-deriv-HO-left-first-1}), where $\beta=1$, we can write Eq.~(\ref{Eq_thermal-deriv-HO1}) as,
\begin{eqnarray}
8  \frac{\partial  \rho T}{\partial t_2} + 8 c u_i \frac{\partial \sigma_{i,j}}{\partial x_{1,j}}
-8c^2 \frac{\partial}{\partial x_{1,j}} \left[ \Pi_{1,i,j} u_i +2  \rho T \tau \frac{\partial  T}{\partial x_{1,j}} \right]  \nonumber \\
- 8 c   \frac{\partial }{\partial x_{1,j}} \left[ \left( \kappa - 2 c  \rho T \tau \right) \frac{\partial T}{\partial x_{1,j}} \right]   
-8 c  \frac{\partial u_i \left( \sigma_{1,i,j} - c \Pi_{1,i,j} \right)}{\partial x_{1,j}}
=0,
\end{eqnarray}
which can be arranged further as,
\begin{eqnarray}
\label{heat_transfer_2nd}
\frac{\partial  \rho T}{\partial t_2}=   c  \sigma_{1,i,j} \frac{\partial u_i}{\partial x_{1,j}}
+  c  \frac{\partial}{\partial x_{1,j}} \left(   \kappa \frac{\partial T}{\partial x_{1,j}}  \right).
\end{eqnarray}
It is  the heat transfer equation in the $\epsilon^2$ order.

In the case of the monoatomic ideal gas at $\beta=0$, Eq.~(\ref{Eq_thermal-deriv-HO1}) can be written as,
\begin{eqnarray}
8  \frac{\partial  \rho T}{\partial t_2} + 8 c u_i \frac{\partial  \Pi_{1,i,j}}{\partial x_{1,j}}
-8c^2 \frac{\partial}{\partial x_{1,j}} \left[ \Pi_{1,i,j} u_i +2 P   \tau \frac{\partial T}{\partial x_{1,j}} \right]=0,
\end{eqnarray}
which can be arranged further as,
\begin{eqnarray}
\frac{\partial  \rho T}{\partial t_2}=   c  \Pi_{1,i,j} \frac{\partial u_i}{\partial x_{1,j}}
+    \frac{\partial}{\partial x_{1,j}} \left(   2 c^2 P  \tau \frac{\partial T}{\partial x_{1,j}}  \right).
\end{eqnarray}
It is equivalent to Eq.~(\ref{heat_transfer_2nd}) if the heat conductivity $\kappa$ is $2c P \tau$ and the shear stress tensor has shear viscosity of $c P \tau$ and zero bulk viscosity. 

Assembling the temporal derivatives up to the $\epsilon^2$ order, one derives the heat-transfer equation,
\begin{eqnarray}
\frac{\partial  \rho T}{\partial t}  + c \frac{\partial u_j  \rho T}{\partial x_{j}} = - c  P \frac{\partial u_j}{\partial x_{j}} +c  \sigma_{i,j} \frac{\partial u_i}{\partial x_{j}}
+  c  \frac{\partial}{\partial x_{j}} \left(   \kappa \frac{\partial T}{\partial x_{j}}  \right),
\end{eqnarray}
with the second-order accuracy of $\epsilon$ at $\beta=1$.
In the case of the ideal monoatomic gas at $\beta=0$, the heat conductivity $\kappa$ is $2c P \tau$ and the shear stress tensor has shear viscosity of $c P \tau$ and zero bulk viscosity.

\section{Derivation of the H-theorem}
\label{appendix_deriv_Hth}

The functional $\mathcal{H}$ defined in Eq.~(\ref{H_fnc}) is analyzed using the non-local equilibrium state in Eq.~(\ref{new_eq_def}) with an assumption of $E_p \le \mathcal{O} \left( \epsilon \cdot \max \left\{ \tilde{\mu}, \tilde{\kappa} \right\} / c \tau \right)$. In addition, $\tilde{\mu} \ge 0$ and $\tilde{\kappa} \ge 0$, which can be realized by adjusting $c$ and $\tau$ if necessary, are assumed.
To show the monotonic behavior of $\mathcal{H}$, after multiplying $\ln f +1$ to the BGK Boltzmann equation, Eq.~(\ref{BGK-Boltzmann}), and taking integration for $v$ and $x$, one can derive,
\begin{eqnarray}
\frac{d \mathcal{H}}{d t} = - \frac{1}{\tau} \int \int \left( f - \tilde{f}^{eq} \right) \left( \ln f - \ln \tilde{f}^{eq} \right) dv dx -  \frac{1}{\tau} \int \int \left( f  - \tilde{f}^{eq} \right) \ln \tilde{f}^{eq} dv dx.
\end{eqnarray}
Due to the monotonicity of the logarithm, the first kernel in the right-hand side is obviously negative. 
If the second kernel is expanded by $\epsilon$ as done in Sec.~\ref{Derivation_of_Navier_Stokes_and_heat_transfer_equations},  we obtain,
\begin{eqnarray}
 \label{eps_expansion_Hth}
 - \int \int \left( f -\tilde{f}^{eq} \right) \ln \tilde{f}^{eq} dv dx 
 =  -  \int \int \left[ \epsilon \left( f^{(1)} -\tilde{f}^{(eq,1)} \right) \ln \tilde{f}^{(eq,0)}  \right. \nonumber \\
\left. - \epsilon^2   \left\{ \left( f^{(2)} -\tilde{f}^{(eq,2)} \right) \ln \tilde{f}^{(eq,0)} +  \left( f^{(1)} -\tilde{f}^{(eq,1)} \right) \frac{\tilde{f}^{(eq,1)}}{\tilde{f}^{(eq,0)}} \right\}  \right] dv dx  + \mathcal{O} \left( \epsilon^3 \right) .
\end{eqnarray}
The term at $\epsilon$ order and the first term at $\epsilon^2$ order in Eq.~(\ref{eps_expansion_Hth}) are zero due to the conservation of mass, momentum, and energy. As a result,
\begin{eqnarray}
\label{eps_expansion_Hth2}
 - \int \int \left( f -\tilde{f}^{eq} \right) \ln \tilde{f}^{eq} dv dx \approx  -  \epsilon^2  \int \int    \left( f^{(1)} -\tilde{f}^{(eq,1)}\right)  \frac{\tilde{f}^{(eq,1)}}{\tilde{f}^{(eq,0)}}  dv dx.
\end{eqnarray}
Using the results of the Chapman-Enskog expansion for the BGK Boltzmann equation, Eq.~(\ref{BGK-Boltzmann}), at $\epsilon$ order,
\begin{eqnarray}
 f^{(1)} -\tilde{f}^{(eq,1)} = - \tau \left( \frac{\partial}{\partial t} + c v_j \frac{\partial }{\partial x_j} \right) \tilde{f}^{(eq,0)},
\end{eqnarray}
it becomes,
\begin{eqnarray}
\label{eps_expansion_Hth3}
\mbox{Eq.~(\ref{eps_expansion_Hth2})} =  \tau  \epsilon^2  \int \int  \left[ \left( \frac{\partial}{\partial t} + c v_j \frac{\partial }{\partial x_j} \right) \tilde{f}^{(eq,0)}  \right] \frac{\tilde{f}^{(eq,1)}}{\tilde{f}^{(eq,0)}}  dv dx.
\end{eqnarray}
Using
\begin{eqnarray}
\frac{\partial \tilde{f}^{(eq,0)}}{\partial t} = \frac{ \tilde{f}^{(eq,0)}}{\rho} \frac{\partial \rho}{\partial t} + \frac{v_l -u_l}{T} \tilde{f}^{(eq,0)} \frac{\partial u_l}{\partial t} + \left( \frac{\left( v - u \right)^2}{2T^2} - \frac{1}{T} \right) \tilde{f}^{(eq,0)} \frac{\partial T}{\partial t}, \\
v_j \frac{\partial \tilde{f}^{(eq,0)}}{\partial x_j} = v_j \frac{ \tilde{f}^{(eq,0)}}{\rho} \frac{\partial \rho}{\partial x_j} + v_j \frac{v_l -u_l}{T} \tilde{f}^{(eq,0)} \frac{\partial u_l}{\partial x_j} + v_j \left( \frac{\left( v - u \right)^2}{2T^2} -  \frac{1}{T} \right) \tilde{f}^{(eq,0)} \frac{\partial T}{\partial x_j},
\end{eqnarray}
one finds the temporal derivative terms and a few spatial derivative terms in Eq.~(\ref{eps_expansion_Hth3}) vanish.
It is because $\int \tilde{f}^{(eq,1)} \vert v \vert^l dv$ where $l \le 2 $ contains only derivative terms related $E_p$ as shown in \ref{deriv_table} and Table.~\ref{tab:moment_new_eq}. Because $E_p \le \mathcal{O} \left( \epsilon \cdot \max \left\{ \tilde{\mu}, \tilde{\kappa} \right\} / c \tau \right)$,  $\int \tilde{f}^{(eq,1)} \vert v \vert^l dv=0$ at $\epsilon$ order.
As a result,
\begin{eqnarray}
\label{eps_expansion_Hth4}
\mbox{Eq.~(\ref{eps_expansion_Hth3})} =c \tau  \epsilon^2  \int \int   v_j  \tilde{f}^{(eq,1)} \left[  \frac{v_l }{T}  \frac{\partial u_l}{\partial x_{1,j}} +  \frac{v^2 - 2 v_l u_l}{2T^2}  \frac{\partial T}{\partial x_{1,j}} \right] dv dx, \nonumber \\
=c \tau  \epsilon^2  \left[ - \frac{\beta c}{T} \tilde{\mu} \tilde{\sigma}_{1,i,j} \frac{\partial u_l}{\partial x_{1,j}} 
-\frac{\beta c \tilde{\kappa}}{T^2} \left| \frac{\partial T}{\partial x_{1,j}} \right|^2  \right], \nonumber \\
=c \tau  \epsilon^2  \left[ - \frac{\beta c}{T} \frac{\tilde{\mu}}{2} \left( \tilde{\sigma}_{1,i,j} \tilde{\sigma}_{1,i,j} + \left| \frac{\partial u_l}{\partial x_{1,l}} \right|^2  \right) 
-\frac{\beta c \tilde{\kappa}}{T^2} \left| \frac{\partial T}{\partial x_{1,j}} \right|^2  \right],
\end{eqnarray}
using
\begin{align}
\int \begin{pmatrix} v_x v_y \\  v^2_x \\ v^3_x \\ v_x v^2_y  \end{pmatrix} \tilde{f}^{(eq,1)} d v_x dv_y=  
\begin{pmatrix} 
- \beta \tilde{\mu} \tilde{\sigma}_{x,y} \\ - \frac{1}{2} \beta \tilde{\mu} \left( \tilde{\sigma}_{x,x} - \tilde{\sigma}_{y,y} \right) \\ -\frac{3}{2} \beta \tilde{\mu} u_x \left( \tilde{\sigma}_{x,x} - \tilde{\sigma}_{y,y} \right) - \frac{3}{2} \beta \tilde{\kappa} \frac{\partial T}{\partial x}
\\ - \beta \tilde{\mu}  \left( u_x \tilde{\sigma}_{y,y} + u_y \left(  \tilde{\sigma}_{x,y} + \tilde{\sigma}_{y,x} \right) \right)  -\frac{1}{2} \beta \tilde{\kappa} \frac{\partial T}{\partial x}
\end{pmatrix}.
\end{align}
As a result, because $0 \le \beta$,  Eq.~(\ref{eps_expansion_Hth3}) should be zero or negative if $\tilde{\mu}$ and $\tilde{\kappa}$ are zero or positive. 
Note that the leading term of the left-hand side in Eq.~(\ref{eps_expansion_Hth2}), Eq.~(\ref{eps_expansion_Hth3}), includes only terms related to $\tilde{\mu}$ and $\tilde{\kappa}$ but does not include the terms related to $E_p$. Since the former terms are always larger than the latter terms because of $E_p \le \mathcal{O} \left( \epsilon \cdot \max \left\{ \tilde{\mu}, \tilde{\kappa} \right\} / c \tau \right)$, it is not necessary to explicitly estimate them here. 
Accordingly Eq.~(\ref{eps_expansion_Hth2}) is zero or negative and therefore $\frac{d \mathcal{H}}{d t} \le 0$ where $E_p \le \mathcal{O} \left( \epsilon \cdot \max \left\{ \tilde{\mu}, \tilde{\kappa} \right\} / c \tau \right)$, $\tilde{\mu} \ge 0$, and $\tilde{\kappa} \ge 0$.

\section{Derivation of the transport coefficients from the fluctuation-dissipation theorem}
\label{appendix_deriv_FDT}

The transport coefficients such as the viscosity and the thermal conductivity are derived from the fluctuation-dissipation theorem. 
In this derivation, we choose an inertial frame so that the local fluid velocity is zero, $u \vert_{x} =0$ and assume $E_p \le \mathcal{O} \left( \epsilon \right)$. 
According to the fluctuation-dissipation theorem, the following integration for the time correlation are considered,
\begin{align}
\label{FDT_mu_RHS}
\frac{c}{T} \int ^{\infty}_{0} \langle P_{xy} \left( t \right)  P_{xy} \left( 0 \right) \rangle dt, \\
\label{FDT_kappa_RHS}
\frac{c}{2  T^2} \int ^{\infty}_{0} \langle \mathcal{S}_i (t) \mathcal{S}_i (0)  \rangle dt,
\end{align}
where $P_{xy}$ is the x-y component of a microscopic pressure tensor and  $\mathcal{S}_i$ is microscopic heat flux.
To obtain the shear viscosity, the first correlation is taken for the flow subject to the shear stress, namely where $\vert \partial_x u_y  \vert\ne 0$.
To obtain the thermal conductivity, the second correlation is taken for the flow subject to the thermal stress, $\vert \partial_x T \vert \ne 0$.

Assuming the state is close to the non-local equilibrium state, one writes Eq.~(\ref{FDT_mu_RHS}) and Eq.~(\ref{FDT_kappa_RHS}) as;
\begin{align}
\label{RHS_FDT1}
 \mbox{Eq.~(\ref{FDT_mu_RHS})}=  \frac{c}{T} \int \int \int ^{\infty}_{0} v_x v_y v'_x v'_y   \tilde{f}^{eq}\left( v \right) \mathcal{P} \left( v', t | v, 0 \right)  dt dv dv', \\
 \label{RHS_FDT_kappa}
 \mbox{Eq.~(\ref{FDT_kappa_RHS})}= \frac{c}{2  T^2}  \int \int \int ^{\infty}_{0}  \left(\frac{v^2_x + v^2_y}{2} - 2 T \right) \left(\frac{v^{\prime 2}_x + v^{\prime 2}_y}{2} - 2 T \right) 
 \left(v_x v^{\prime}_x + v_y v^{\prime}_y \right)  \nonumber \\  \tilde{f}^{eq}\left( v \right) \mathcal{P} \left( v', t | v, 0 \right)  dt dv dv'.
\end{align}
 Here $\mathcal{P} \left( v', t | v, 0 \right)$ is the conditional probability of finding the particle of velocity $v'$ at time $t$ where the particle of velocity $v$ was observed at time 0. It may be written with the functional derivatives as,
\begin{align}
\label{conditiona_probab_expand}
\mathcal{P} \left( v', t | v, 0 \right) 
= \frac{\delta f \left( v', t\right)}{\delta f \left( v, 0\right)}
= \frac{\delta f \vert_{v', t}}{\delta f \vert_{v, 0}} 
+ \int^{t}_{0} \int \frac{\delta f \vert_{v', t}}{\delta \tilde{f}^{eq} \vert_{v'', t'}}  \frac{\delta \tilde{f}^{eq} \vert_{v'', t'}}{\delta f \vert_{v, 0}} dv'' dt'  + \cdots. 
\end{align}
The first term in the right-hand side shows the correlation between an initial state and a state after advection;
\begin{align}
 \frac{\delta f \vert_{v', t}}{\delta f \vert_{v, 0}} = \exp \left( - \frac{t}{\tau} \right) \delta \left( v' - v \right) .
\end{align}
The exponential decay is due to the collision during the particle's advection.
The second term in the right-hand side of Eq.~(\ref{conditiona_probab_expand}) shows the correlation between an initial state and an advected state after a collision. 
Specifically, particles collide after advecting from an initial state with velocity $v$ and then follow the equilibrium state. After that, the equilibrium state advects back to the original position at time $t$ with velocity $v'$ which is the opposed number to $v$.
It may be further written as,
\begin{align}
\label{conditional_prob_secondkernel}
\int^{t}_{0} \int \frac{\delta f \vert_{v', t}}{\delta \tilde{f}^{eq} \vert_{v'', t'}}  \frac{\delta \tilde{f}^{eq} \vert_{v'', t'}}{\delta f \vert_{v, 0}} dv'' dt'  
=\int^{t}_{0} \int   \exp \left( - \frac{t-t'}{\tau} \right) \delta \left( v' - v'' \right)   \frac{\delta \tilde{f}^{eq} \vert_{v'', t'}}{\delta X \vert_{t'}} \frac{ \delta X \vert_{t'}}{\delta f \vert_{ v, 0 }} \delta \left( t'- \tau \right)  dv'' dt', \nonumber \\
=\int  \exp \left( - \frac{t-t'}{\tau} \right) \frac{\delta \tilde{f}^{eq} \vert_{v', t'}}{\delta X \vert_{t'}} \frac{ \delta X \vert_{t'}}{\delta f \vert_{ v, 0 }} \delta \left( t'- \tau \right)   dt',
\end{align}
where  $X= \left\{ \rho, u_{\alpha}, T, \nabla_{\alpha} u_{\beta}, \nabla_{\alpha} T \right\}$, the macroscopic quantities defining the equilibrium state $\tilde{f}^{eq}$, and $\alpha$ and $\beta$ are indices for the spatial direction.
In the second equation, to simplify the time integration for $t'$, the collision is assumed to happen at time $\tau$, which results in the Dirac delta function $\delta \left( t'- \tau \right) $. 

Next, the functional derivative $\frac{ \delta X \vert_{t'}}{\delta f \vert_{ v, 0 }}$ is estimated. 
With the definition of the functional derivative,
\begin{align}
\label{func_deriv_def}
\int \frac{ \delta X }{\delta f} \left( v \right) \phi \left( v \right) dv = \lim_{\epsilon \rightarrow 0} \frac{X \left[ f + \epsilon \phi \right]- X \left[ f \right]}{\epsilon},
\end{align}
where $\phi$ is an arbitrary function, one can write the case of $X= \rho$ as,
\begin{align}
\label{func_deriv_rho}
\int \frac{ \delta \rho \vert_{x,t} }{\delta f \vert_{v,x,0}} \phi \left( v \right) dv = \lim_{\epsilon \rightarrow 0} \frac{\rho \left[ f + \epsilon \phi \vert_{x,0} \right] \vert_{x,t} - \rho \left[ f \right]  \vert_{x,t}  }{\epsilon}.
\end{align}
Since the function $\phi$ takes the values along the direction of $f \vert_{v,x,0}$, the first term of the numerator in the right-hand side can be written as,
\begin{align}
\rho \left[ f + \epsilon \phi \vert_{x,0} \right] = \int \left\{  f (v) \vert_{x,t} + \epsilon \phi \mathcal{F}(v) \vert_{x,t} \right\} dv, \\
\mathcal{F}(v) \vert_{x,t}= \delta\left( t \right) + \delta \left( v \right) \exp \left( - \frac{t}{\tau}\right). 
\end{align}
The first term in $\mathcal{F}(v) $ is a correlation between $f \vert_{v,x,0}$ and $f \vert_{v,x,t}$ at time $0$ and the second term is a correlation between $f \vert_{v,x,0}$ and $f \vert_{v,x,t}$ at $v=0$.
Applying it to Eq.~(\ref{func_deriv_rho}), one derives,
\begin{align}
\frac{ \delta \rho \vert_{x,t} }{\delta f \vert_{v,x,0}} = \delta\left( t \right) + \delta \left( v \right) \exp \left( - \frac{t}{\tau}\right) .
\end{align}

Similarly, in the case of $X = u_{\alpha}$, because
\begin{align}
u_{\alpha} \left[ f + \epsilon \phi \vert_{x,0} \right] 
=  \frac{   \int \left\{  f\vert_{x,t} + \epsilon \phi \mathcal{F}(v) \vert_{x,t} \right\} v_{\alpha} dv }{  \int \left\{  f\vert_{x,t} + \epsilon \phi \mathcal{F}(v) \vert_{x,t} \right\} dv}, \nonumber \\
=\frac{\int f v_{\alpha} dv}{\int f dv} 
+ \epsilon \frac{\int \phi \mathcal{F}(v) \vert_{x,t}   v_{\alpha} dv}{\rho}
- \epsilon \frac{\rho u_{\alpha}}{\rho^2} \int \mathcal{F}(v) \vert_{x,t}  dv + \mathcal{O} \left( \epsilon^2 \right),
\end{align}  
using the Taylor expansion, $\frac{A + \epsilon}{B+ \epsilon^{\prime}}=\frac{A}{B}+ \epsilon \frac{1}{B}- \epsilon^{\prime} \frac{A}{B^2} + \cdots$, one obtains,
\begin{align}
\frac{ \delta u_{\alpha} \vert_{x,t} }{\delta f \vert_{v,x,0}} = \frac{v_{\alpha} }{\rho} \delta \left( t \right),
\end{align}
if $u_{\alpha}\vert_{x}=0$ is considered.

In the case of $X =T$, using the definition of $T=  \int f v^2 dv /\left( 2 \rho \right)$ where $u \vert_{x} =0$ and the Taylor expansion, $\frac{A + \epsilon}{B+ \epsilon^{\prime}}=\frac{A}{B}+ \epsilon \frac{1}{B}- \epsilon^{\prime} \frac{A}{B^2} + \cdots$, because
\begin{align}
T \left[ f + \epsilon \phi \vert_{x,0} \right] =  T \left[ f \right] + \epsilon \left\{ 
\frac{\int \phi \mathcal{F}(v) \vert_{x,t} v^2 dv}{2 \rho }
-\frac{\int \phi \left\{ \mathcal{F}(v) \vert_{x,t} 2 \rho T  dv \right\} }{2 \rho^2  } \right\}+ \mathcal{O} \left( \epsilon^2 \right), \nonumber \\
=  T \left[ f \right] + \epsilon \int \phi \mathcal{F}(v) \left( \frac{\rho v^2 - 2 \rho T}{2 \rho^2} \right) dv + \mathcal{O} \left( \epsilon^2 \right), \nonumber \\
=  T \left[ f \right] + \epsilon \int \phi \left\{ \frac{v^2 - 2 T}{2 \rho}  \left( \delta \left( t \right) + \delta \left( v \right) \exp \left( -\frac{t}{\tau}\right) \right) \right\} dv + \mathcal{O} \left( \epsilon^2 \right),
\end{align}
one obtains,
\begin{align}
\frac{\delta T \vert_{x,t}}{\delta f \vert_{v,x,0}} =   \frac{v^2 - 2 T}{2 \rho}  \left( \delta \left( t \right) + \delta \left( v \right) \exp \left( -\frac{t}{\tau}\right) \right).
\end{align}

In the case of $X =\partial_{\alpha} u_{\beta}$, using $\partial_{\alpha} u_{\beta} \vert_{x,t} \approx \left( \int f \vert_{x,t} v_{\beta} dv  - \int f\vert_{x - \Delta x_{\alpha},t}  v_{\beta} dv  \right) /  \Delta x_{\alpha}$, because
\begin{align}
\partial_{\alpha} u_{\beta} \vert_{x,t} \left[ f + \epsilon \phi \vert_{x,0} \right] =\partial_{\alpha} u_{\beta} \vert_{x,t} \left[ f \right] 
- \epsilon \int \phi \frac{v_{\beta}}{ \rho \Delta x_{\alpha}} \exp \left( -\frac{t}{\tau} \right) \delta \left(v_{\alpha} t c 
- \Delta x_{\alpha} \right)  dv 
+ \mathcal{O} \left( \epsilon^2 \right),
\end{align}
if $u_{\alpha}\vert_{x}=0$ is considered, one obtains,
\begin{align}
\label{Func_deriv_gradu}
\frac{\delta \partial_{\alpha} u_{\beta}  \vert_{x,t}}{\delta f \vert_{v,x,0}} = -\frac{v_{\beta}}{\rho v_{\alpha} t c}  \exp \left( -\frac{t}{\tau} \right).
\end{align}

In the case of $X =\partial_{\alpha} T$, using $\partial_{\alpha} T \vert_{x,t} \approx \left\{ \left(  \int f v^2 dv /2 \rho \right) \vert_{x,t}  - \left( \int f v^2 dv  / 2 \rho \right) \vert_{x-\Delta x_{\alpha} ,t}  \right\}/  \Delta x_{\alpha}$ where $u \vert_{x} =0$, because
\begin{align}
\partial_{\alpha} T\vert_{x,t}  \left[ f + \epsilon \phi \vert_{x,0} \right] =\partial_{\alpha} T \vert_{x,t}  \left[ f \right] 
- \epsilon \int \phi \frac{1}{  \Delta x_{\alpha}} \frac{v^2 - 2 T }{2 \rho }\exp \left( -\frac{t}{\tau} \right) \delta \left(v_{\alpha} t c - \Delta x_{\alpha} \right)  dv 
+ \mathcal{O} \left( \epsilon^2 \right),
\end{align}
one obtains,
\begin{align}
\label{Func_deriv_gradT}
\frac{\delta \partial_{\alpha} T  \vert_{x,t}}{\delta f \vert_{v,x,0}} = -\frac{1}{  v_{\alpha} t c} \frac{v^2 - 2 T }{2 \rho} \exp \left( -\frac{t}{\tau} \right).
\end{align}

In summary, in the inertial frame of $u \vert_{x} =0$, the functional derivative for the macroscopic quantities $X$ can be written as,
\begin{align}
 \label{X_Func_deriv_sum}
 \frac{\delta X \vert_{t'}}{\delta f \vert_{v,0}} 
 = \left\{ \delta\left( t' \right) + \delta \left( v \right) \exp \left( - \frac{t'}{\tau}\right),
  \frac{v_{\alpha} }{\rho} \delta \left( t' \right) ,  \right. \nonumber \\
\left. \frac{v^2 - 2 T}{2 \rho}  \left( \delta \left( t' \right) + \delta \left( v \right) \exp \left( -\frac{t'}{\tau}\right) \right), 
 -\frac{v_{\beta}}{\rho v_{\alpha} t' c} \exp \left( -\frac{t'}{\tau} \right),
 -\frac{1}{  v_{\alpha} t' c} \frac{v^2 - 2 T }{2 \rho  }\exp \left( -\frac{t'}{\tau} \right)
  \right\}.
\end{align}

\hspace{5cm}

If we consider applying Eq.~(\ref{X_Func_deriv_sum}) to Eq.~(\ref{conditional_prob_secondkernel}) , we find that all terms related to $\delta \left( t' \right)$ in Eq.~(\ref{X_Func_deriv_sum}) are negligible due to the delta function, $\delta \left( t'- \tau \right) $ in Eq.~(\ref{conditional_prob_secondkernel}).
In addition, if we consider applying Eq.~(\ref{conditiona_probab_expand}) and Eq.~(\ref{conditional_prob_secondkernel}) to Eq.~(\ref{RHS_FDT1}), and Eq.~(\ref{RHS_FDT_kappa}) using Eq.~(\ref{X_Func_deriv_sum}), we find that all terms related to $\delta \left( v \right)$ in Eq.~(\ref{X_Func_deriv_sum})  are negligible because both of $P_{xy}$ and  $\mathcal{S}_i$ are zero if particle velocity is zero.
As a result, we should consider only the derivatives of Eq.~(\ref{Func_deriv_gradu}) and Eq.~(\ref{Func_deriv_gradT}).

First, for the flow under the shear stress, 
using Eq.~(\ref{Func_deriv_gradu}), Eq.~(\ref{conditional_prob_secondkernel}) can be written as,
\begin{align}
\int^{t}_{0} \int \frac{\delta f \vert_{v', t}}{\delta \tilde{f}^{eq} \vert_{v'', t'}}  \frac{\delta \tilde{f}^{eq} \vert_{v'', t'}}{\delta f \vert_{v, 0}} dv'' dt'  
=\int  \exp \left( - \frac{t-t'}{\tau} \right) \beta f^{eq} \frac{\tilde{\mu} }{\rho T^2 }  v^{\prime}_{\alpha} v^{\prime}_{\beta}    \frac{v_{\beta}}{\rho v_{\alpha} t' c} \exp \left( -\frac{t'}{\tau} \right) \delta \left( t'- \tau \right)   dt', \nonumber \\
= \exp \left( - \frac{t}{\tau} \right) \beta f^{eq} \frac{\tilde{\mu} }{\rho T^2 }  v^{\prime}_{\alpha} v^{\prime}_{\beta}    \frac{v_{\beta}}{\rho v_{\alpha} \tau c}.
\end{align}
Here the following equation derived from Eq.~(\ref{new_eq_def}) and Eq.~(\ref{delta_feq}) was used,
\begin{align}
\frac{\delta \tilde{f}^{eq} \vert_{v', t'}}{\delta \partial_{\alpha} u_{\beta} \vert_{t'}} =
-\beta f^{eq} \vert_{v', t'} \frac{\tilde{\mu} }{\rho T^2 }  v^{\prime}_{\alpha} v^{\prime}_{\beta}. 
\end{align}
As a result, the conditional probability, Eq.~(\ref{conditiona_probab_expand}), can be written as,
\begin{align}
\mathcal{P} \left( v', t | v, 0 \right) 
= \exp \left( - \frac{t}{\tau} \right) \delta \left( v' - v \right) 
+ \exp \left( - \frac{t}{\tau} \right) \beta f^{eq} \frac{\tilde{\mu} }{\rho T^2 }  v^{\prime}_{\alpha} v^{\prime}_{\beta}    \frac{v_{\beta}}{\rho v_{\alpha} \tau c} + \cdots. 
\end{align}
If it is substituted to Eq.~(\ref{RHS_FDT1}), one obtains,
\begin{align}
 \mbox{Eq.~(\ref{FDT_mu_RHS})} 
 \approx  \frac{c}{  T} \int \int \int ^{\infty}_{0} v_x v_y v'_x v'_y   \tilde{f}^{eq}\left( v \right)  \left\{\exp \left( - \frac{t}{\tau} \right) \delta \left( v' - v \right)  \right. \nonumber \\
\left. + \exp \left( - \frac{t}{\tau} \right) \beta f^{eq} \frac{\tilde{\mu} }{\rho T^2 }  v^{\prime}_{\alpha} v^{\prime}_{\beta}    \frac{v_{\beta}}{\rho v_{\alpha} \tau c} \right\}dt dv dv',  \nonumber \\
= \frac{c}{  T} \left( \tau \int v^2_x v^2_y \tilde{f}^{eq}\left( v \right)  dv + \frac{\tilde{\mu} }{\rho T^2 c} \int \int v^{\prime 2}_{x} v^{\prime 2}_{y}    \frac{v^2_{y}}{\rho } \tilde{f}^{eq}\left( v \right) f^{eq}\left( v' \right)  dv dv' \right)
\approx \tau \rho T c + \tilde{\mu} \approx \mu.
\end{align}
Here $E_p \le \mathcal{O} \left( \epsilon \right)$ and the following equations were used, 
\begin{align}
\label{feq_moment_u0}
\int \begin{pmatrix} 1 \\  v^2_x \\ v^4_x \\ v^6_x \\ v^2_x v^2_y \end{pmatrix} f^{eq} d v_x dv_y=  
\begin{pmatrix} 
\rho \\  \rho T  \\  3 \rho T^2  \\ 15 \rho T^3  \\ \rho T^2 
\end{pmatrix},
 \end{align}
 where $\left| u \right|=0$.
 
 For the flow under the thermal stress, using Eq.~(\ref{Func_deriv_gradT}), Eq.~(\ref{conditional_prob_secondkernel}) can be written as,
\begin{align}
\int^{t}_{0} \int \frac{\delta f \vert_{v', t}}{\delta \tilde{f}^{eq} \vert_{v'', t'}}  \frac{\delta \tilde{f}^{eq} \vert_{v'', t'}}{\delta f \vert_{v, 0}} dv'' dt'  \nonumber \\
=\int^{t}_{0}    \exp \left( - \frac{t-t'}{\tau} \right) \beta f^{eq} \left( v' \right) \frac{\tilde{\kappa} }{2 \rho T^3 }  \left( v^{\prime 2} - 4  T \right) v^{\prime}_{\alpha} \frac{1}{  v_{\alpha} t' c} \frac{v^2 - 2 T }{2 \rho } \exp \left( -\frac{t'}{\tau} \right) \delta \left( t' - \tau \right)   dt', \nonumber \\
= \exp \left( - \frac{t}{\tau} \right) \beta f^{eq} \left( v' \right)  \frac{\tilde{\kappa} }{2 \rho T^3 }  \left( v^{\prime 2} - 4  T \right) v^{\prime}_{\alpha} \frac{1}{  v_{\alpha} \tau c} \frac{v^2 - 2 T }{2 \rho }.  \nonumber \\
\end{align}
Here the following equation derived from Eq.~(\ref{new_eq_def}) and Eq.~(\ref{delta_feq}),
\begin{align}
\frac{\delta \tilde{f}^{eq} \vert_{v', t'}}{\delta \partial_{\alpha} T \vert_{t'}} =
-\beta f^{eq} \vert_{v', t'} \frac{\tilde{\kappa} }{4 \rho T^3 }  \left( v^{\prime 2} - 4  T \right) v^{\prime}_{\alpha},
\end{align}
was used.
As a result, the conditional probability, Eq.~(\ref{conditiona_probab_expand}), can be written as,
\begin{align}
\mathcal{P} \left( v', t | v, 0 \right) 
= \exp \left( - \frac{t}{\tau} \right) \delta \left( v' - v \right) 
+ \exp \left( - \frac{t}{\tau} \right) \beta f^{eq} \left( v' \right)  \frac{\tilde{\kappa} }{2 \rho T^3 }  \left( v^{\prime 2} - 4  T \right) v^{\prime}_{\alpha} \frac{1}{  v_{\alpha} \tau c} \frac{v^2 - 2 T }{2 \rho }  + \cdots. 
\end{align}
If it is substituted to Eq.~(\ref{RHS_FDT_kappa}), using Eq.~(\ref{feq_moment_u0}) one obtains,
\begin{align}
\mbox{Eq.~(\ref{FDT_kappa_RHS})}= \nonumber \\
\frac{c}{2  T^2} \int \int \int^{\infty}_{0} \left( \frac{v^2_x + v^2_y}{2} - 2  T \right)  \left( \frac{v^{\prime 2}_x + v^{\prime 2}_y}{2} - 2  T \right) \left( v_x v^{\prime}_x + v_y v^{\prime}_y \right) \tilde{f}^{eq} \left( v \right)  \mathcal{P} \left( v', t | v, 0 \right)  dt dv dv', \nonumber \\
= \frac{c}{2  T^2} \left\{ \tau \int  \left( \frac{v^2_x + v^2_y}{2} - 2  T \right)^2   \left( v^2_x  + v^2_y \right) \tilde{f}^{eq} \left( v \right)   dv \right. \nonumber \\
\left.  + \beta   \frac{\tilde{\kappa} }{4 \rho^2 T^3 c}
\int \int   \left( \frac{v^2_x + v^2_y}{2} - 2  T \right)  \tilde{f}^{eq} \left( v \right)   \frac{v^2 - 2 T}{  v_{\alpha}} \left( v_x v^{\prime}_x + v_y v^{\prime}_y \right)   \right. \nonumber \\
\left.  \left( \frac{v^{\prime 2}_x + v^{\prime 2}_y}{2} - 2  T \right) f^{eq} \left( v^{\prime} \right) \left( v^{\prime 2} - 4  T \right) v^{\prime}_{\alpha} dv dv'
\right\}, \nonumber \\
\approx  \frac{\tau c}{2 T^2  } 4 \rho T^3  +  \frac{1}{2 T^2  } \frac{\tilde{\kappa} }{4 \rho^2 T^3 } \left( 2 \rho T^2  \right) \left( 4 \rho T^3  \right)
= 2 \rho T  \tau c+ \tilde{\kappa} \approx \kappa,
\end{align}
where Eq.~(\ref{feq_moment_u0}) and  $E_p \le \mathcal{O} \left( \epsilon \right)$ are used.

\section{Derivation of Eq.~(\ref{BGK-Ceicigani2})}
\label{deriv_FPterm}

The non-local equilibrium state, Eq.~(\ref{new_eq_def}), is applied to the drifting and diffusion terms in the Fokker-Planck equation,
\begin{align}
\label{eq_appendix_derivFPterm}
 \frac{\partial^2}{\partial v^2} \left( \bar{D}  \tilde{f}^{eq} \right)+ \frac{\partial}{\partial v_i} \left\{ \left(v_i-u_i \right) \gamma \tilde{f}^{eq} \right\},
\end{align}
with assumptions of incompressible limit $\partial u_i / \partial x_i =0$, $\bar{D}=\gamma T$, $E_p=0$, and negligible temperature variation.
With the assumptions, we can regard $\tilde{u}=u$ and $\tilde{T}=T$ in Eq.~(\ref{u_renormalization}) and Eq.~(\ref{T_renormalization}). 
For convenience,  $g^{eq}$ is defined as followings,
\begin{align}
g^{eq} = \frac{\tilde{\mu} }{\rho T^2 } \left\{ \left(v_l -u_l \right) \left(v_k -u_k \right)  - \frac{\left( v-u \right)^2}{2} \delta_{l,k} \right\} \frac{\partial u_l}{\partial x_k}.
\end{align}
Therefore,
\begin{equation}
\tilde{f}^{eq} = f^{eq} + \delta f^{eq}= f^{eq} - \beta  f^{eq} g^{eq}.
\end{equation}
Then Eq.~(\ref{eq_appendix_derivFPterm}) can be written as,
\begin{align}
\mbox{Eq.~(\ref{eq_appendix_derivFPterm})}=
g^{eq} \left\{ \bar{D}  \frac{\partial^2 f^{eq}}{\partial v^2} + \gamma f^{eq} + \gamma \left( v_i - u_i \right)  \frac{\partial f^{eq}}{\partial v_i} \right\} \nonumber \\
- \beta \bar{D} \left( 2 \frac{\partial f^{eq}}{\partial v_i} \frac{\partial g^{eq}}{\partial v_i} + f^{eq} \frac{\partial^2 g^{eq}}{\partial v^2}  \right) 
-\beta \gamma f^{eq} g^{eq} 
- \beta \gamma \left( v_i - u_i \right)  f^{eq} \frac{\partial g^{eq}}{\partial v_i}.
\end{align}
As well known in the Fokker-Planck equation, the terms in the first parentheis disapper using $\bar{D}=\gamma T$ and the following relationships,
\begin{align}
 \frac{\partial f^{eq}}{\partial v_i} = - \frac{v_i - u_i}{T} f^{eq}, \nonumber \\
 \frac{\partial^2 f^{eq}}{\partial v^2} = - \frac{f^{eq}}{T} + \left( \frac{v-u}{T} \right)^2 f^{eq}.
\end{align}
For the rest of terms, using,
\begin{align}
 \frac{\partial g^{eq}}{\partial v_i} = \frac{\tilde{\mu} }{\rho T^2 } \left\{ \delta_{i,l} \left(v_k -u_k \right) +\delta_{i,k} \left(v_l -u_l \right)  - \left( v_i -u_i \right) \delta_{l,k} \right\} \frac{\partial u_l}{\partial x_k}, \nonumber \\
 \left( v_i - u_i \right)   \frac{\partial g^{eq}}{\partial v_i} = 2 g^{eq}, \nonumber \\
  \frac{\partial^2 g^{eq}}{\partial v^2}  = \frac{\tilde{\mu} }{\rho T^2 } \frac{\partial u_l}{\partial x_l}=0,
\end{align}
we can obtain,
\begin{align}
\mbox{Eq.~(\ref{eq_appendix_derivFPterm})}=
2 \beta \gamma f^{eq} g^{eq}= -2 \gamma \delta f^{eq}.
\end{align}

\end{document}